\documentclass[iop]{emulateapj}
\usepackage{here}
\usepackage{amsmath}

\shorttitle{Evolution of Blister-Type HII Regions in a Magnetized Medium}
\shortauthors{Gendelev \& Krumholz}

\begin{document}

\slugcomment{Accepted by The Astrophysical Journal}
 
\title{Evolution of Blister-Type HII Regions in a Magnetized Medium}

\author{Leo Gendelev\altaffilmark{1} and Mark R. Krumholz \altaffilmark{1}}

\affil{Department of Astronomy, University of California, Santa Cruz, CA 95064}

\begin{abstract}

We use the three-dimensional Athena ionizing radiation-magnetohydrodynamics (IRMHD) code to simulate blister-type HII regions driven by stars on the edge of magnetized gas clouds. We compare these to simulations of spherical HII regions where the star is embedded deep within a cloud, and to non-magnetized simulations of both types, in order to compare their ability to drive turbulence and influence star formation. We find that magnetized blister HII regions can be very efficient at injecting energy into clouds. This is partly a magnetic effect: the magnetic energy added to a cloud by an HII region is comparable to or larger than the kinetic energy, and magnetic fields can also help collimate the ejected gas, increasing its energy yield. As a result of these effects, a blister HII region expanding into a cloud with a magnetic field perpendicular to its edge injects twice as much energy by 5 Myr as a non-magnetized blister HII region driven by a star of the same luminosity. Blister HII regions are also more efficient at injecting kinetic energy than spherical HII regions, due to the recoil provided by escaping gas, but not by as much as predicted by some analytic approximations. 
\end{abstract}

\keywords{HII regions; ISM: clouds; ISM: kinematics and dynamics; ISM: magnetic fields; MHD; radiative transfer; stars: formation}

\section{Introduction}

 In recent years it has become evident that HII regions -- formed when massive stars ionize their surroundings -- are of crucial importance to the evolution of giant molecular clouds (GMCs) since they photoevaporate their nascent clouds, trigger star formation, and drive turbulence. \citet{McKeeWilliams1997a} find that massive stars inject much more energy into the interstellar medium (ISM) by emitting ionizing radiation than through supernovae. Furthermore, semi-analytic models show that HII regions can disrupt GMCs entirely, through a combination of photoevaporation and mechanical disruption, over a period of a few cloud crossing times -- roughly 20 - 40 Myr \citep{McKeeWilliams1997b,Matzner2002,KrumholzMatznerMckee2006,Goldbaum2011}. As the ionization front moves through the cloud, it sweeps up neutral gas, potentially increasing the star formation rate in the dense shell (\citet{Elmegreen1977, Whitworth1994}; however see \citet{Dale2007a}, who argues that this effect is generally small.) HII regions may also be important for driving turbulence in GMCs. Observed GMC lifetimes are significantly longer than the free-fall time scale associated with the dissipation of turbulence \citep{Krumholz2007b, Fukui2009}. \citet{Matzner2002} finds that HII regions provide more energy for turbulence than the combined effects of stellar winds and supernovae, and \citet{KrumholzMatznerMckee2006} and \citet{Goldbaum2011} argue that this energy is sufficient to drive turbulence over observed GMC lifetimes. Hence it is important to study HII regions and their energy injection mechanism in greater detail. 
 
 Recent numerical studies have examined HII regions in turbulent media as a possible mechanism to explain the qualitative features and star formation rates observed in GMCs (e.g. \citealt{Mellema2006}, \citealt{Dale2007b}, and \citealt{Gritschneder2009}). These papers show that the interaction of the HII region with the pre-existing turbulent gas has important effects. In \citet{Gritschneder2009} the turbulent energy is significantly increased (up to a factor of 4 in the cold gas) as the ionization heats the gas along channels of low density, compressing higher density gas into filaments with gravitational collapse occuring in the tips of the pillar like structures.

 \citet{Krumholz2007} perform the first ever numerical study of the expansion of an HII region into a magnetized gas. It is important to study the effects of magnetic fields since the magnetic energy in GMCs is comparable to the kinetic and gravitational energies \citep{Crutcher1999}. Over time the ionization front slows down enough due to the resistance by the magnetic field lines in the perpendicular direction that the fast magnetosonic wave outruns it, disturbing neutral gas ahead of the ionization front, so that there is swept up material in between the fast magnetosonic wave and the ionization front. In this region the stretched magnetic field acts as an energy reservoir. 
  
\begin{figure}
  \begin{center}
    \includegraphics[scale=0.40]{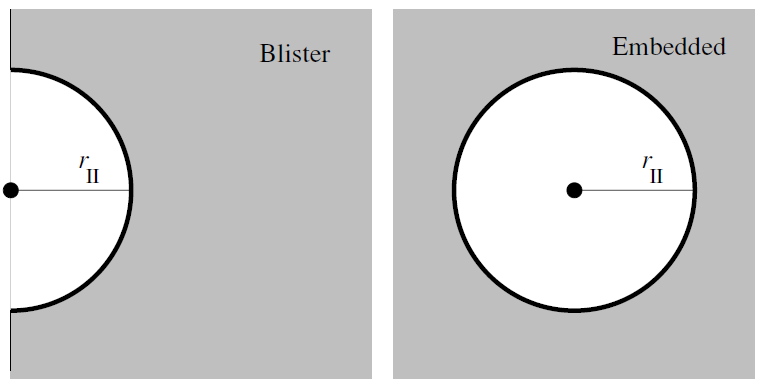}
  \end{center}
  \caption{A sketch comparing the blister and symmetric HII regions \citep{Matzner2009}.}
  \label{fig:blister-symmetric-sketch}
\end{figure}

 In this paper we expand on this work by examining magnetized blister-type HII regions, also known as champagne flows, which form when an ionizing star is situated towards the edge of the GMC. A sketch adopted from \citet{Matzner2009} is shown in Fig. \ref{fig:blister-symmetric-sketch} comparing blister and symmetric HII regions.  In the blister case the star is situated next to the edge of the GMC, so the ionization front will eventually reach the edge of the cloud and burst a hole through which hot ionized gas will be able to stream out at supersonic velocities into the low-density ISM. The blister scenario was first envisioned by \cite{Whitworth1978} as a mechanism to disperse molecular gas clouds and was studied by \citet{Tenorio-Tagle1978} to explain some features of observed nebulae. In contrast with the symmetric case, the ionized gas within the HII region is not confined to the HII region, leading to an increased expansion rate and kinetic energy of the ionization front due to the ``rocket effect'' \citep{Kahn1954}. In the symmetric case the gas is confined within the HII region and cannot rocket away, so this effect does not apply. Hence we would expect more energy to be injected into the GMC by the blister HII region. It might seem coincidential that a star would be positioned close to the edge of a molecular gas cloud, but observational evidence points to this being the rule rather than the exception \citep{Israel1978}. This makes sense given that GMCs are turbulent -- the turbulence creates a filamentary structure in the GMC so that any star has a high probability of being born near an ``edge''. There are only a few examples of blister HII regions for which the magnetic field can be measured. One is M17, in which \citet{Pellegrini2007} find that the magnetic field is strong enough that it has essentially halted the expansion of the HII region.

 Since \citet{Krumholz2007} there have been several numerical studies to take MHD effects into account. \citet{Henney2009} and \citet{Mackey2010} performed IRMHD (ionizing radiation magnetohydrodynamical) simulations of magnetized globules. Given strong initial magnetic fields, \citet{Henney2009} observed substantial deviations from symmetry, in particular when the initial magnetc field was oriented perpendicular to the direction of ionizing radiation. The ionized gas as a whole was found to be dominated by magnetic pressure, and evidence was found that magnetic effects might be important in the formation of bright, bar like emission features in HII regions, a result confirmed by \citet{Mackey2010}. Even more recently \citet{Arthur2011} performed simulations of HII regions expanding into a magnetized gas with turbulence, resulting in a morphology of striking similarity to observed HII regions with clearly destinguishable pillar like structures. However, none of these studies have dealt with blister HII regions, and it would be difficult for those of them that focus on turbulent regions to do so, because these simulations use periodic boundary conditions which cannot account for gas being ``blown off'' the computational grid -- and in a turbulent medium there is always the opportunity for material to be blown off. These studies are useful in that they provide evidence for the importance of magnetic fields and the ability of turbulence to reproduce observed features of HII regions, but they need to be followed up by the examination of the energetics rather than the qualitative consequences of MHD turbulence and a comparison of the energy injection efficiency of the symmetric and blister-type HII regions. 
 
 Our work is the first numerical study of blister type HII regions evolving in the presence of uniform magnetic fields but without turbulence and is therefore complementary to previous work on symmetric HII regions in a turbulent medium with and without MHD. The structure for the rest of the paper is as follows: in \S \ref{sec:computational-methodology} we present our computational approach and parameters, in \S \ref{section:results} we present our results from the various types of runs, looking at symmetric and blister-type HII regions both with magnetic fields of varying orientations and without magnetic fields, in \S \ref{sec:discussion} we discuss the results and compare to analytic approximations, and in \S \ref{section:conclusion} we draw our conclusions. 
 
\section{Computational Methodology}

\label{sec:computational-methodology}

We perform the simulations with the Athena 3D MHD code \citep{Stone2008}. Athena is a grid-based, static mesh refinement (SMR) code designed for simulating various astrophysical MHD processes. It uses higher order Godunov methods, which are particularly efficient when used with static or adaptive mesh refinement, in combination with the constrained transport technique, which is used to ensure that the magnetic divergence is preserved to machine precision. In addition, we employ the radiation scheme first introduced in \citet{Krumholz2007}. We run all simulations on the Pleiades cluster at UCSC at $256^{3}$ resolution using 64 processors for an average wall-time of 3 days per simulation.

\subsection{The Ideal MHD Equations}

Athena solves the equations of ideal MHD. In conservative form, they are:

\begin{eqnarray}
	 \label{eqn:continuity}
  \frac{\partial \rho}{\partial t} + \nabla \cdot(\rho \textit{\textbf{v}})&=& 0 \\
   \label{eqn:momentum}
    \frac{\partial}{\partial t}(\rho \textit{\textbf{v}}) + \nabla \cdot(\rho \textit{\textbf{vv}} - \textit{\textbf{BB}})+\nabla P^{*}  &=& 0 \\
     \label{eqn:faraday}
  \frac{\partial \textit{\textbf{B}}}{\partial t} + \nabla \cdot(\textit{\textbf{vB}} - \textit{\textbf{Bv}}) &=& 0 \\
     \label{eqn:energy}
  \frac{\partial E}{\partial t} + \nabla \cdot [(E + P^{*})\textit{\textbf{v}}
  - \textit{\textbf{B}}
  (\textit{\textbf{B}} \cdot \textit{\textbf{v}})] &=& \mathcal{G}-\mathcal{L} \\
  \label{eqn:chemistry}
  \frac{\partial \rho_{n}}{\partial t} + \nabla \cdot(\rho_{n} \textit{\textbf{v}})&=& \mathcal{R} - \mathcal{I} \\
  \label{eqn:bfield}
  \nabla \cdot \textit{\textbf{B}} &=& 0,
\end{eqnarray}
where $\rho$ is the density and $\textit{\textbf{v}}$ is the velocity of the gas, $\textit{\textbf{B}}$ is the magnetic field, $P^{*}=P+(\textit{\textbf{B}}\cdot \textit{\textbf{B}}/2)$ is the total pressure, $P$ is the gas thermal pressure, $E$ is the total energy density, and $\rho_{n}$ is the density of the neutral gas. Equation \eqref{eqn:continuity} is the continuity equation (conservation of mass), \eqref{eqn:momentum} is the conservation of momentum, where we have used the approximation that our fluid has no viscosity ($\nu = 0$), \eqref{eqn:faraday} is Faraday's law, or the induction equation, where we have set the magnetic diffusivity to zero ($\eta = 0$), \eqref{eqn:energy} is the conservation of energy equation, where $\mathcal{G}$ and $\mathcal{L}$ are the radiative heating and cooling terms, respectively, \eqref{eqn:chemistry} is the continuity equation for neutral gas, and says that a change in the mass of neutral gas can only come through advection into other cells or through recombinations ($\mathcal{R}$) and ionizations ($\mathcal{I}$), and \eqref{eqn:bfield} comes from the non-existence of magnetic monopoles. 

\subsection{Heating, Cooling, and Ionization Chemistry}

All heating, cooling, and ionization terms are as in \cite{Krumholz2007}, where the heating and cooling rates are adopted from \citet{Koyama2002} for neutral gas, and from \citet{Osterbrock1989} for partially ionized gas.  The heating and cooling equations are provided below:

\begin{equation}
\mathcal{G}= e_{\gamma}\sigma n_{H}\mathop{\sum}_{n}\frac{s_{n}}{4\pi |\textbf{x}-\textbf{x}_{n}|^{2}}e^{-\tau(\textbf{x},\textbf{x}_{n})} + n_{H}\Gamma_{KI},
\end{equation}

\begin{eqnarray}
\mathcal{L}= \Lambda _{rec}n_{e}n_{H^{+}}+\Lambda_{ff}(T)n_{e}n_{H^{+}}+\Lambda_{KI}(T)n_{H}^{2} \nonumber \\
+ \Lambda_{line}(T) n_{e}n_{H^{+}},
\end{eqnarray} where $n_H$, $n_{H^+}$, and $n_e$ are the number densities of neutral hydrogen, ionized hydrogen, and electrons, respectively, $e_\gamma$ is the thermal energy added per ionization, $\sigma$ is the cross section for absorption of a photon at the ionization threshold for neutral hydrogen, $s_n$ and $\mathbf{x}_n$ are the ionizing photon luminosity and position of the $n$th star, $\mathbf{x}$ is the position where the ionization is taking place, $\tau$ is the optical depth to ionizing photons between $\mathbf{x}$, and $\mathbf{x}_n$, given by 
\begin{equation}
\tau(\textbf{x},\textbf{x}_{n})=\int\limits_{\textbf{x}_{n}}^{\textbf{x}} \!(\sigma n_{H})\, \mathrm{d}l.
\end{equation} $\Lambda_{rec}$, $\Lambda_{ff}$, and $\Lambda_{line}$ are the cooling rates due to recombination radiation, free-free emission, and metal line emission in the ionized gas, and $\Gamma_{KI}$ and $\Lambda_{KI}$ are the neutral ISM heating and cooling rates, computed using the approximation of \citet{Koyama2002}.

We should point out a particular feature of heating and cooling rates that apply in the neutral medium (where $n_{H^+}=0$): these particular heating and cooling curves have the property that there is a two phase equilibrium. At a given pressure there are two solutions: a high temperature - low density solution, and a low temperature - high density solution. We use this property of the neutral gas to initially set up our blister type problems, by placing the high density and low density halves of the computational domain in pressure equilbrium. 

\subsection{Problem Setup}

\label{sec:problem-setup}

We set up a rectangular grid which runs from -25.0 to 25.0 pc in all 3 directions. The computational domain has outflow boundary conditions. In all runs we place a star at the center of the grid, with $(x,y,z)=(0,0,0)$. We then use two possible initial density distributions. In the spherical case, we initialize the grid to a uniform initial density $n=63.0$ H atoms cm$^{-3}$ across the whole domain. For blister-type runs, we split the grid into two halves: all cells with $x < 0$ are of higher density, $n_{\rm left} = 63.0$ cm$^{-3}$, while all cells with $x > 0$ are of a low density, $n_{\rm right} = 0.055$ cm$^{-3}$. Given our cooling curves, the equilibrium temperature at a density of $n=63.0$ cm$^{-3}$ is $T = 55.0$ K (sound speed $c_0 = 5.74 \times 10^4$ cm s$^{-1}$), while at a density of $n=0.055$ cm$^{-3}$ it is $T = 6.3 \times 10^{3}$ K (sound speed $c_0 = 6.14 \times 10^5$ cm s$^{-1}$). Thus the two sides are in pressure balance. The sound speed inside the HII region is $c_{i}=8 \times 10^{5}$. The mean particle mass for the neutral gas is $2.3 \times 10^{-24}$ grams. For the spherical MHD and the blister MHD runs we thread the initial magnetic field through the domain in the $\hat{x}$ direction. In addition, we perform one blister-type run with the initial magnetic field at $45^{\circ}$ to the $x$ axis and finally one run with the magnetic field in the $\hat{y}$ direction. In all MHD runs, the initial magnetic field is $B_{0} = 3.0 \times 10^{-6}$ in the code units, which differ from cgs units by a factor of $\sqrt{4\pi}$, so that in cgs it is 10.6$\mu G$. A summary of some of these parameters is provided in Tab. \ref{table:problem-setup}.

\begin{deluxetable*}{llllllll}
\tabletypesize{\footnotesize}
\tablecolumns{8} 
\tablewidth{0pt} 
\tablecaption{Problem Setup \label{table:problem-setup}}
\tablehead{
\colhead{ Simulation Name}  & \colhead{Type} & \colhead{B-Field Setup} & \colhead{B-Field} & \colhead{$n_{left}$} & \colhead{$n_{right}$} & \colhead{$T_{left}$} & \colhead{$T_{right}$}
}
\startdata
Hydro & Symmetric & None & None & 63.0 cm$^{-3}$ & 63.0 cm$^{-3}$ & 55.0 K & 55.0 K \\
MHD & Symmetric & $\hat{x}$ direction & 10.6 $\mu G$ & 63.0 cm$^{-3}$ & 63.0 cm$^{-3}$ & 55.0 K & 55.0 K\\
Blister-hydro & Blister & None & None & 63.0 cm$^{-3}$ & 0.055 cm$^{-3}$ & 55.0 K & 6.3 $\times 10^{3}$ K \\
Blister-mhd & Blister &  $\hat{x}$ direction & 10.6 $\mu G$ & 63.0 cm$^{-3}$ & 0.055 cm$^{-3}$ & 55.0 K & 6.3 $\times 10^{3}$ K\\
Blister-mhd-vert & Blister & $\hat{y}$ direction  & 10.6 $\mu G$ & 63.0 cm$^{-3}$ & 0.055 cm$^{-3}$ & 55.0 K & 6.3 $\times 10^{3}$ K\\
Blister-mhd-45 & Blister & $45^{\circ}$  & 10.6 $\mu G$ & 63.0 cm$^{-3}$ & 0.055 cm$^{-3}$ & 55.0 K & 6.3 $\times 10^{3}$ K\\ 
\enddata
\end{deluxetable*}

We parameterize the luminosity of our central source in terms of its Stromgren radius \citep{Stromgren1939}, which is defined as:
\begin{equation}
r_{s}=\left (\frac{3s\mu_{H}^2}{4\pi \alpha^{(B)}\rho^{2}} \right )^{1/3},
\label{eqn:Stromgren-Radius}
\end{equation}
where $s$ is the ionizing luminosity of the star, $\mu_H$ and $\rho$ are the mean mass per hydrogen atom and the density of the gas, respectively, and $\alpha^{(B)}$ is the recombination coefficient. We set $r_s = 1.5$ pc for al the simulations, computed using the value of $\rho$ for the dense half of the grid in the blister cases. The corresponding ionizing luminosity is $s = 5.3\times 10^{47}$ s$^{-1}$, appropriate for a star of spectral type B0.5.

We configure Athena with the Roe solver based on the Godunov scheme in conjunction with the ctu integrator to produce the most accurate results, and we enable h-correction in order to eliminate carbuncle problems. We resort to using first order fluxes because higher order fluxes proved unstable at $256^3$ resolution in at least some of our runs. All other parameters relevent to our problem required to configure Athena so as to reproduce our results are in \citet{Krumholz2007}. 

A note on the blister type HII region setup: in reality it would clearly be a coincidence if the star was positioned right on the edge of the cloud. Blister-type HII regions are likely to form when a star is close to, but not directly on, the edge of the cloud. However ours is an instructive limiting case, since it is much harder to interpret and make sense of computational data that would result from a more realistic setup. It would be useful to extract as much information as possible from our idealized setup, and in a later paper expand our investigation to compare to the more realistic scenario such as M17 \citep{Pellegrini2007}). 

\section{Results}                 
\label{section:results}
\subsection{Symmetric Simulations}
\label{sec:symmetric-simulations}

We begin our analysis by revisiting the extensively studied classic - the symmetric HII region - where the star is situated deep within the cloud. 

\subsubsection{Hydro Run}
\label{sec:hydro-run}
\begin{figure*}
  \begin{center}
    \includegraphics[scale=0.33]{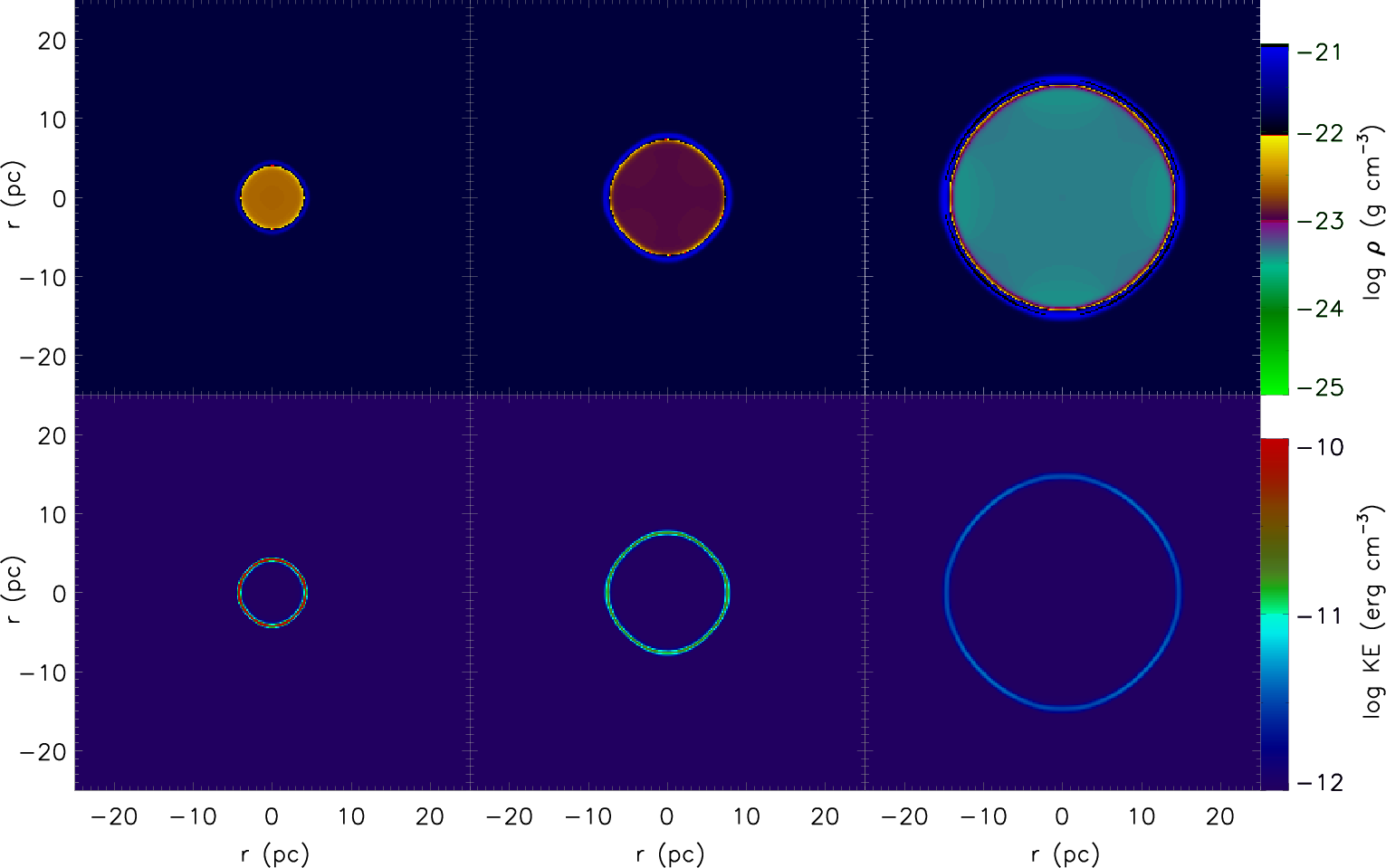}
  \end{center}
\caption{Slices in the $z=0$ plane taken from the hydro run. The $1^{st}$, $2^{nd}$, and $3^{rd}$ columns correspond to 0.5, 1.5, and 5.0 Myr into the simulation, respectively. The $1^{st}$ and $2^{nd}$ rows display the density and kinetic energy density of the hydro run, respectively.}
\label{fig:hydro-plot}
\end{figure*}

As expected, in the absence of a magnetic field, the ionization front expands in a spherically symmetric shell as shown in Fig. \ref{fig:hydro-plot}. We can see that the density of the shell increases over time and that virtually all the kinetic energy is contained within the thin shell that bounds the ionization front. 

\subsubsection{MHD Run}
\label{sec:mhd-run}
\begin{figure*}
  \begin{center}
    \includegraphics[scale=0.33]{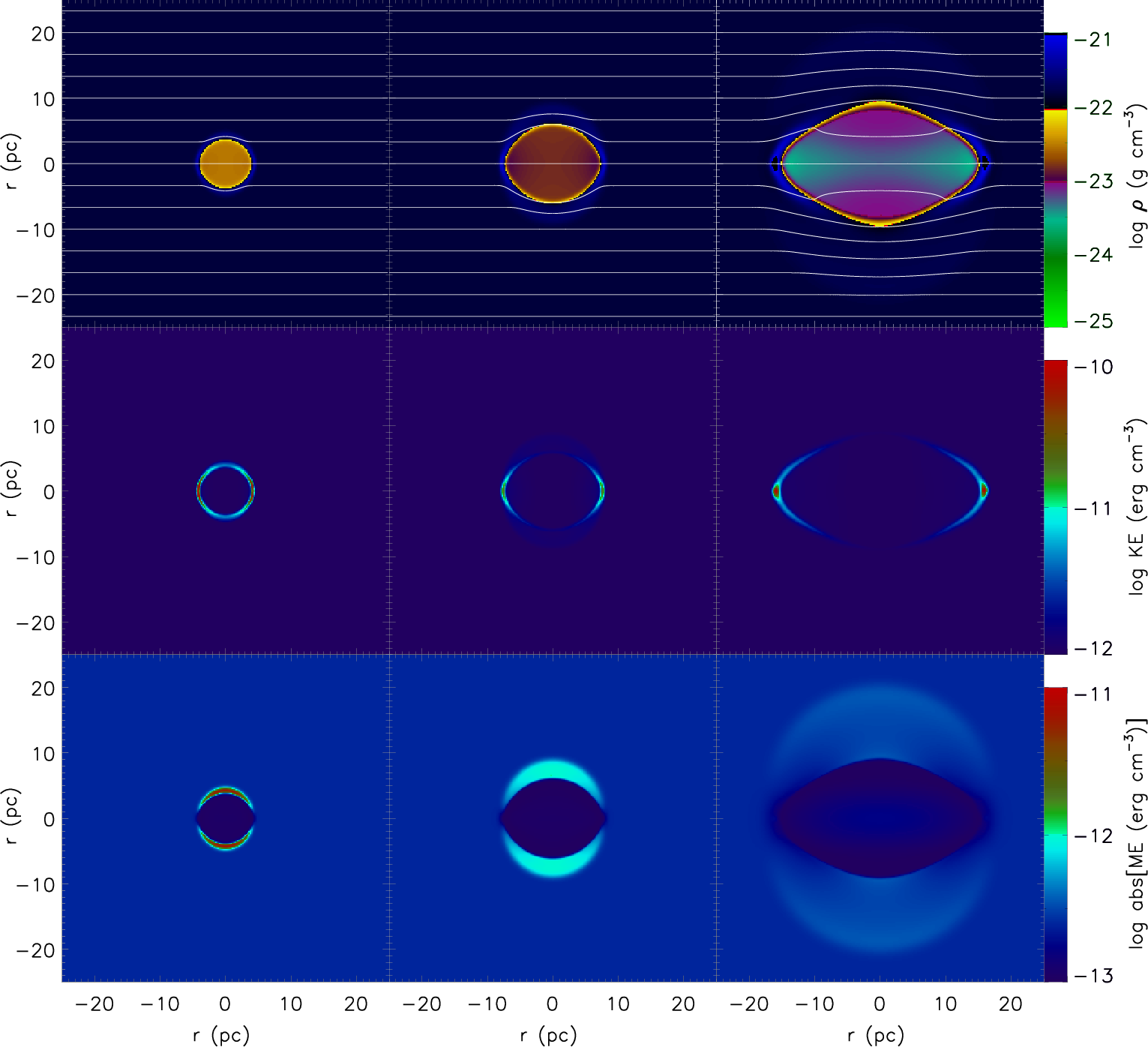}
  \end{center}
\caption{The mhd run - same as Fig. \ref{fig:hydro-plot} except for the addition of the third row, which shows the change of the magnetic energy over time, $\Delta E_{B}$, where $\Delta E_{B} = ((B_{x})^{2}+(B_{y})^{2}+(B_{z})^{2})/2 - ((B_{x_{0}})^{2}+(B_{y_{0}})^{2}+(B_{z_{0}})^{2})/2$, where $B_{x_{0}}, B_{y_{0}}, B_{z_{0}}$ are the initial magnetic field strengths in the x, y, and z directions, respectively. Note that within the HII region interior $\Delta E_{B} < 0$, which is why we plot the log of its absolute value. Everywhere else in the computational domain $\Delta E_{B} > 0$. The white lines in the top row represent the magnetic field lines, drawn from footpoints equally spaced along the left side of the region shown.}
\label{fig:mhd-plot}
\end{figure*}

In the presence of the magnetic field, the expansion of the ionization front is strongly suppressed perpendicular to the magnetic field lines, so that over time the symmetric HII region assumes the shape of a football, as seen in Fig. \ref{fig:mhd-plot}. From row 3, it is apparent that at 0.5 Myr - corresponding to the first column of the figure - the fast magnetosonic wave is just beginning to take the lead in front of the shell. By 1.5 Myr this is one of the most prominent features of the plot.  Most of the added magnetic energy is contained in this region, implying that the magnetic energy injected into the cloud by the HII region increases at late times. 

\subsection{Blister-Type Simulations}
\label{sec:blister-runs}
\subsubsection{Blister-hydro Run}
First we look at the blister-type scenario where there is no magnetic field present, in order to be able to better understand what effects the addition of a magnetic field has on the HII region. 

\begin{figure*}
  \begin{center}
    \includegraphics[scale=0.33]{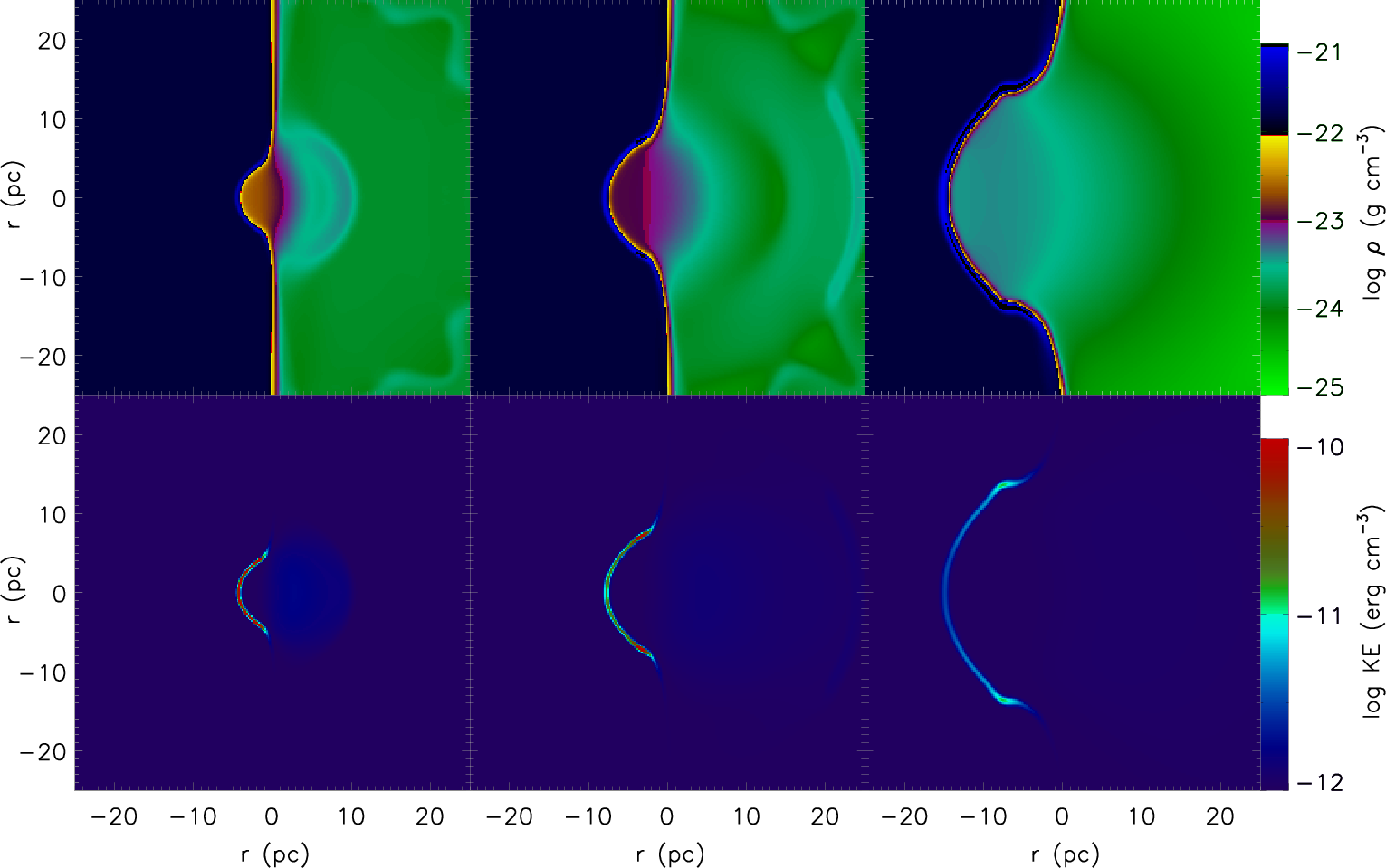}
  \end{center}
\caption{The blister-hydro run: the $1^{st}$ row are the density slices while the $2^{nd}$ row contains the kinetic energy density. The density is plotted on a log scale so that effects in the low-density part of the computational domain can be distinguished. }
\label{fig:blister-hydro}
\end{figure*}

Density and kinetic energy slices of the computational grid are presented for 0.5, 1.5, and 5 Myr of this run in Fig. \ref{fig:blister-hydro}. Initially the expansion into the dense half resembles the expansion in the symmetric case -- the ionization front shell is almost identical to the left hemisphere of the spherical shell in the symmetric non-mhd simulation (Fig. \ref{fig:hydro-plot}). Over time, however, the deviation from symmetry becomes increasingly apparent. By 1.5 Myr there are slivers of the dense shell that extend further in the $\hat{y}$ direction. This effect is easiest to see from the kinetic energy plot in row 2 of the same figure. Unlike the embedded case, there is a jet of gas blowing out of the cloud. Although this low-density material covers a wider area than the dense shell, its average kinetic energy density is orders of magnitude lower than the kinetic energy density within the shell, so its kinetic energy is virtually negligible.  

\subsubsection{Blister-mhd Run}
\label{sec:blister-mhd-run}
\begin{figure*}
  \begin{center}
    \includegraphics[scale=0.33]{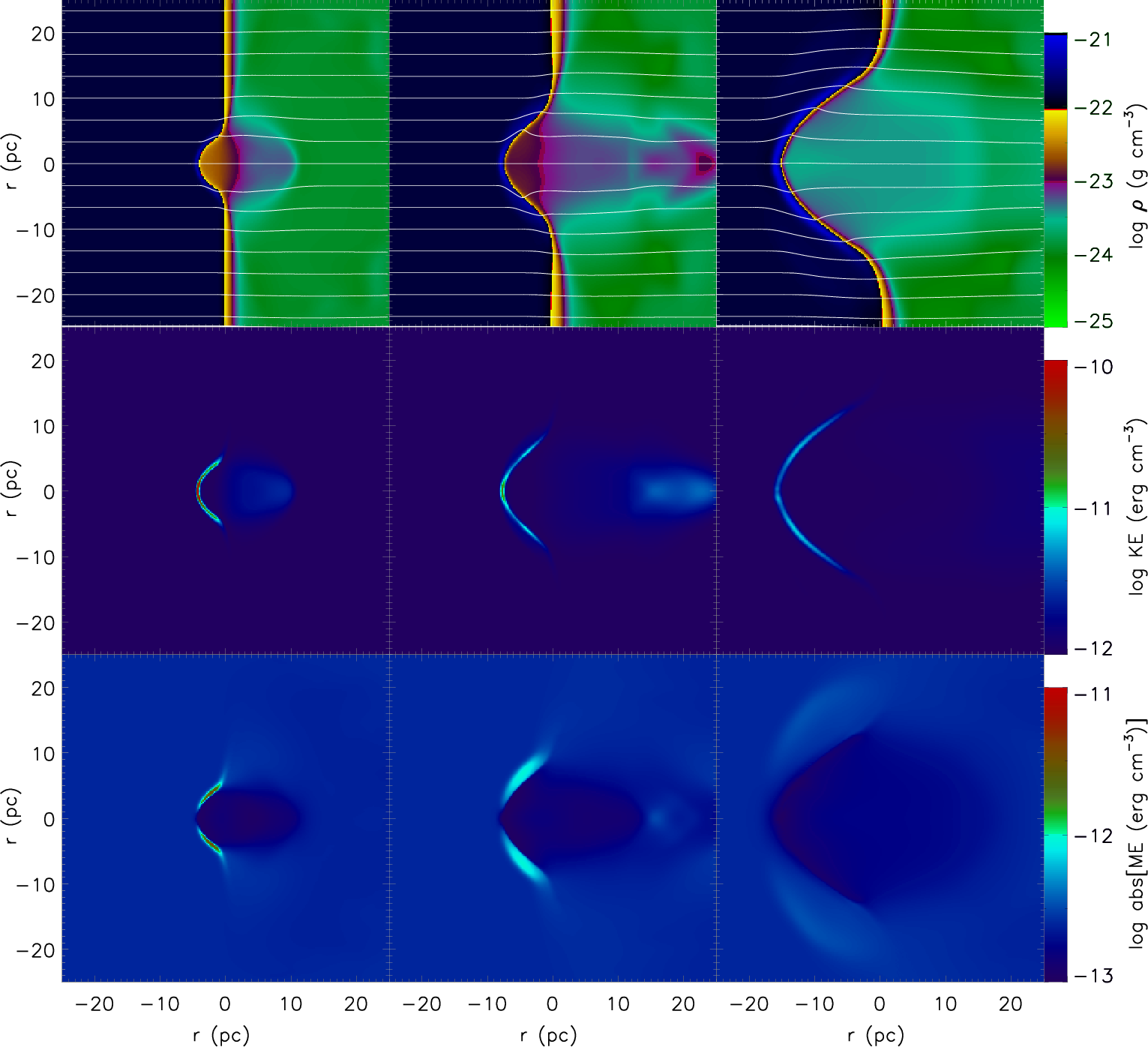}
  \end{center}
\caption{The blister-mhd run. All panels are the same as in Fig. \ref{fig:mhd-plot}.}
\label{fig:blister-mhd}
\end{figure*}

We now present the blister-mhd results that are the focus of this paper. The magnetic field orientation we look at first is particularly useful since it is easy to compare and contrast with the blister-hydro run. The expansion of gas -- both in the dense and low density portions of the computational domain -- is suppressed in the directions perpendicular to the magnetic field (Fig. \ref{fig:blister-mhd}), and the magnetic field streamlines the ionized gas blowing out of the cloud.

A chunk of hot gas resembling a bullet bursts out of the cloud and by 1.5 Myr has already reached the edge of the grid, implying a speed of 15-20 km/s, so a significant amount of kinetic energy is both gained and lost over the course of the simulation. 

The kinetic energy in the dense half is all concentrated in the shell, which is very similar in structure to the shell in the blister-hydro case (row 2 of Fig \ref{fig:blister-mhd}). However, the shell is more oblate, and in contrast to the blister-hydro shell, the kinetic energy decreases much more slowly with time. Thus one of the most important MHD effects in the blister-mhd case is that the magnetic field changes the nature of the expansion over time (as we will see in \S \ref{sec:comparison}), by collimating the jet of gas streaming out of the cloud.

\subsubsection{Blister-mhd-vert Run}

\label{sec:blister-mhd-vert-run}

\begin{figure*}
  \begin{center}
    \includegraphics[scale=0.33]{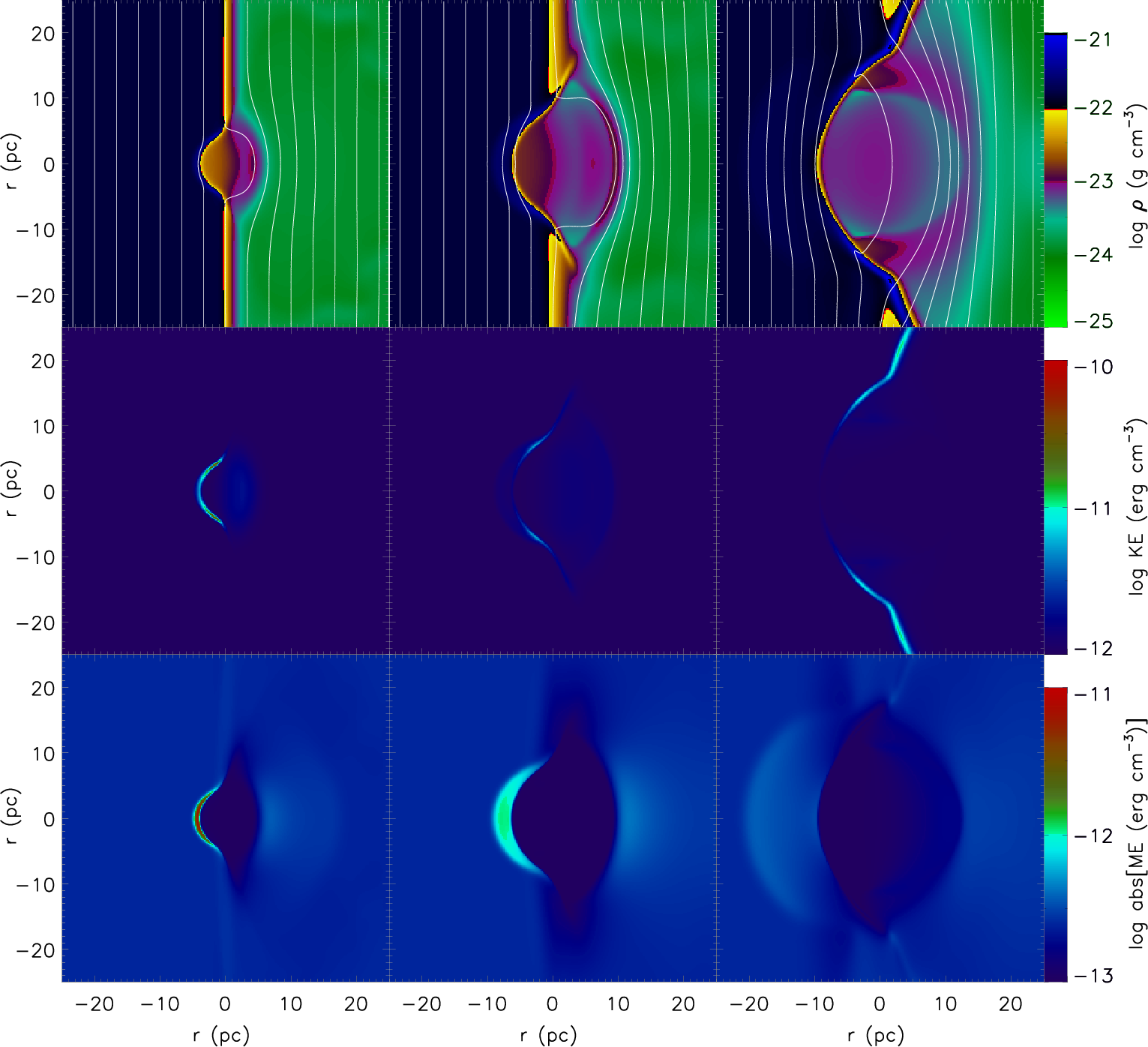}
  \end{center}
\caption{The blister-vert run. All panels are the same as in Fig. \ref{fig:mhd-plot}.}
\label{fig:blister-mhd-vert}
\end{figure*}

Initially, similar to the blister-mhd case, a jet of gas bursts out of the cloud as seen in Fig. \ref{fig:blister-mhd-vert}. However the jet is not streamlined and is not propelled at high velocities through the low-density medium. Its motion is highly suppressed by the magnetic field perpendicular to it. The magnetic field gets stretched the most out of any of the runs.

The kinetic energy of the cloud is concentrated in the shell just as for the other runs, but the magnetic field's suppression of the shell's motion in all but the vertical direction transfers most of the kinetic energy from the spherical part of the shell to the slivers of dense gas protruding into the low density medium by the end of the run ($2^{nd}$ row). The magnetic energy density is plotted in row 3 of Fig. \ref{fig:blister-mhd-vert}. As expected, since the ionization front is slow to expand perpendicular to the magnetic field lines, the fast magnetosonic wave disconnects from the front earlier than in any other simulation, resulting in a very prominent density fluctuation leading the ionization front.

\subsubsection{Blister-mhd-45 Run}

\label{sec:blister-mhd-45-run}

\begin{figure*}
  \begin{center}
    \includegraphics[scale=0.33]{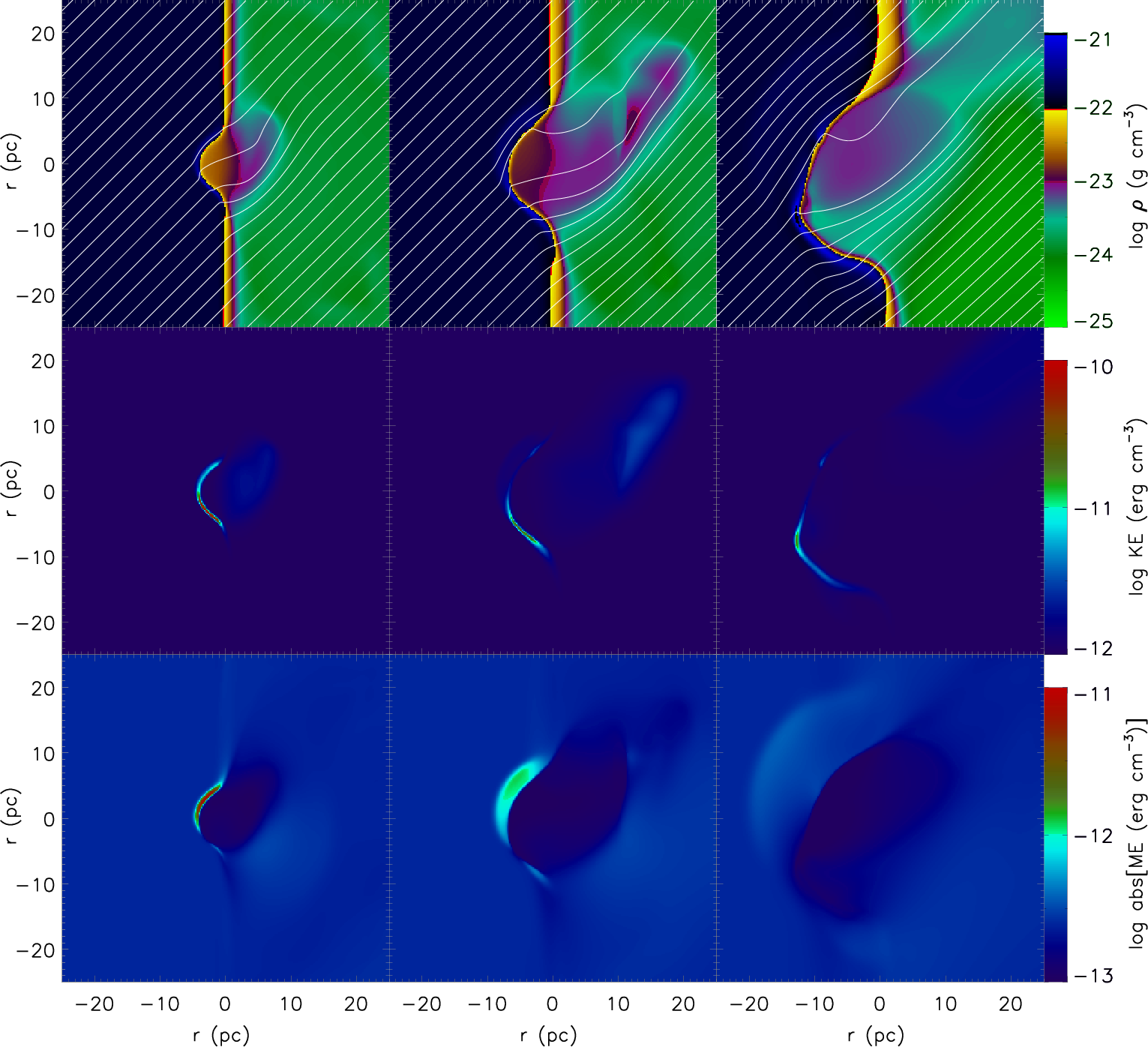}
  \end{center}
\caption{The blister-mhd-45 run. All the panels are the same as in Fig. \ref{fig:mhd-plot}.}
\label{fig:blister-mhd-45}
\end{figure*}

This simulation is clearly a blend of the properties of the blister-mhd and blister-vert runs. As seen in Fig. \ref{fig:blister-mhd-45}, the magnetic field lines limit the expansion of the front perpendicular to them, but do not prevent the front from travelling at a sizeable velocity parallel to them. The result is an HII region that has rectangular structure.

\subsection{Comparison of the Symmetric and Blister Simulations}
\label{sec:comparison}
In this section we compare properties such as the radius, kinetic, magnetic, and total energies of the various simulations.

\subsubsection{Shell Radius and Mass}
\label{sec:expansion-radius}

The radius of the shell as a function of time can be estimated analytically using conservation of momentum and some simplifying assumptions. If we assume that the pressure inside the HII region is vastly dominant over the ambient pressure in the neutral gas into which it is expanding at all times, and that the density within the HII region is approximately uniform, then we can obtain an equation of motion for the shell from momentum conservation. This equation of motion can then be solved by using a power-law similarity solution. We provide a derivation for both the symmetric and blister cases in Appendix \ref{sec:appendix}. In \citet{Spitzer1978} there is a derivation of the symmetric case performed in a slightly different way, yielding a very similar solution. The solutions we find for the two cases differ only by a factor of $2^{2/7}$, and are provided below:
\begin{equation}
\label{Eqn:Spitzer1978-A}
r_{sh}=r_{s}\left ( \frac{7t}{\sqrt{12}t_{s}} \right )^{4/7}       \mbox{(spherical)},
\end{equation}
and
\begin{equation}
\label{Eqn:Spitzer1978-B}
r_{sh}=r_{s}\left ( \frac{7t}{\sqrt6 t_{s}} \right )^{4/7}       \mbox{(blister)},
\end{equation}
where $t_{s}=r_{s}/c_{ii}$. 

How valid are the approximations we used to derive the shell radius (Eqn.'s \eqref{Eqn:Spitzer1978-A} and \eqref{Eqn:Spitzer1978-B})? In reality, even though the assumption that the ambient pressure is negligible in comparison with the pressure inside the HII region is quite accurate at early times, the HII region internal pressure decreases as the HII region expands, and so it becomes less accurate. Furthermore, even though the assumption that the density is uniform within the HII region for the symmetric case is valid, it is not necessarily quite as true for the blister case where gas is free to stream out of the HII region into the interstellar medium. Hence the radius as a function of time will likely not match the analytic solution perfectly for all times, especially for the blister case. This is discussed in more detail in \S \ref{sec:discussion}.

We define the shell radius as the average radius of all cells whose density is greater than or equal to $1.01\rho_0$ (where $\rho_{0}$ is the initial density of the neutral gas) in order to avoid taking into account the neutral, undisturbed gas and the low density medium, as well as to take into account the contribution to the radius from the gas in the fast magnetosonic wave which can have a density of $\rho_0 < \rho < 1.1 \rho_0$. If we were to use a cut-off of say, $1.1 \rho_0$, it would make little difference for the hydro runs, since they are characterized by very thin, high-density shells. For the MHD, runs, though, using a cut-off of $1.1\rho_0$ would not only produce a smaller radius, it would lead us to severely underestimate the added magnetic energy, since most of this energy is contained in the mildly overdense region between the ionization front and the fast magnetosonic wave.

\begin{figure}
\plotone{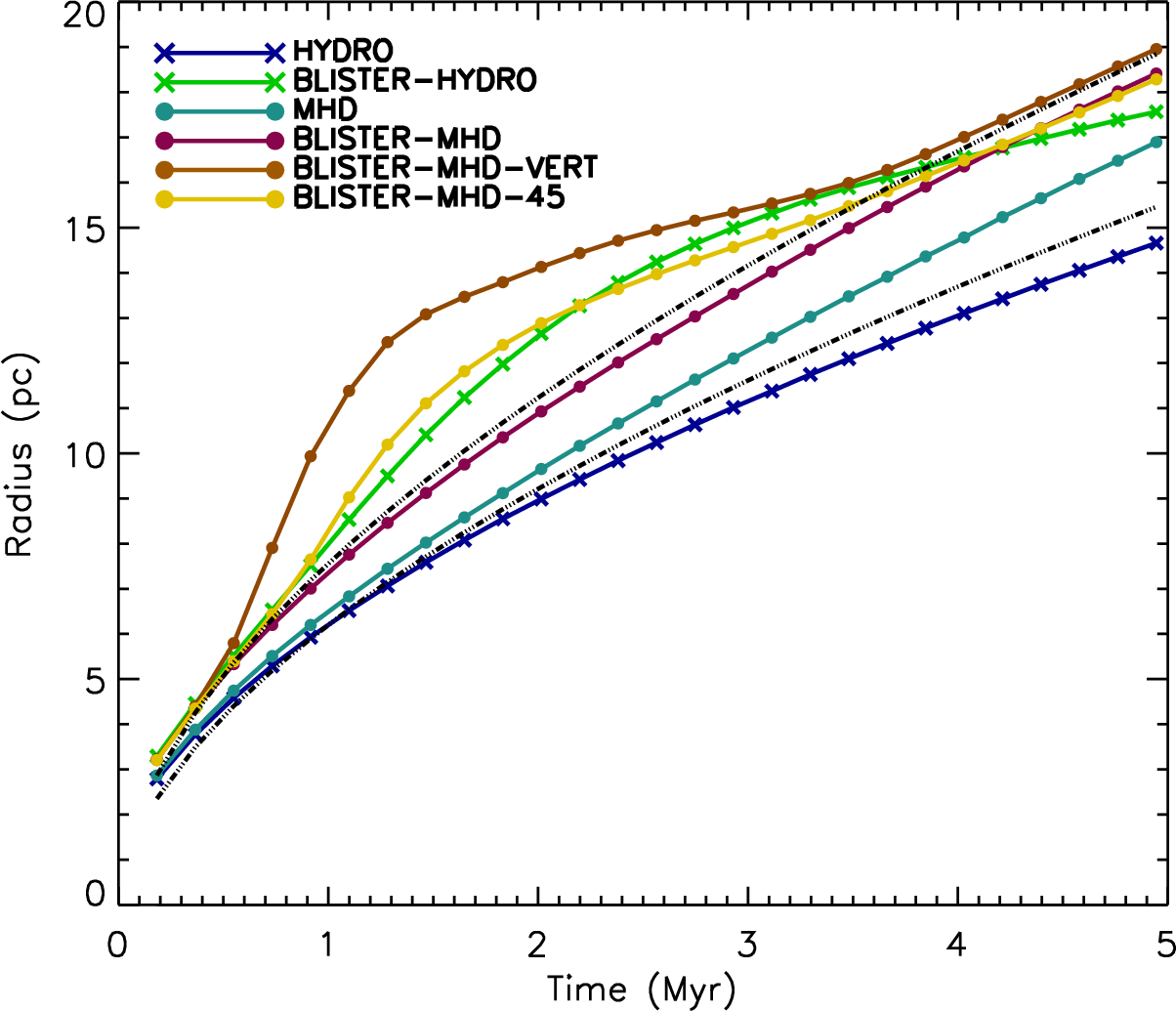}
\caption{The mass averaged radii of all the runs. The blue-dashed curves represent the analytic solutions for the blister (upper curve) and spherical (lower curve) cases respectively.}
\label{fig:radius-comparison} 
\end{figure}

The shell expands slower for the hydro case than the analytic approximation (Eqn. \eqref{Eqn:Spitzer1978-A}). The average radius in the mhd run is greater than in the hydro run due to the fast magnetosonic wave leading the ionization front (Fig.  \ref{fig:radius-comparison}). The radius of the shell is defined using the same criteria as in the hydro run, but this time the definition encompasses the material contained in between the ionization front and the fast magnetosonic wave since this material has a higher density than the neutral background. As expected, all the blister-type runs expand faster than their symmetric counterparts. Although there is some variation between the magnetized blister runs in the early-mid stages of the simulations, at late times they all converge with one another. This is due to the fact that, at late times, the radius is effectively set by the fast magnetosonic fluctuation leading the ionization front. Thus in all the MHD cases, the radius is determined not only by the speed of the actual ionization front, which varies between the cases, but also by the fast magnetosonic speed, which does not. 

\begin{figure}
\centering
\includegraphics[scale=0.19]{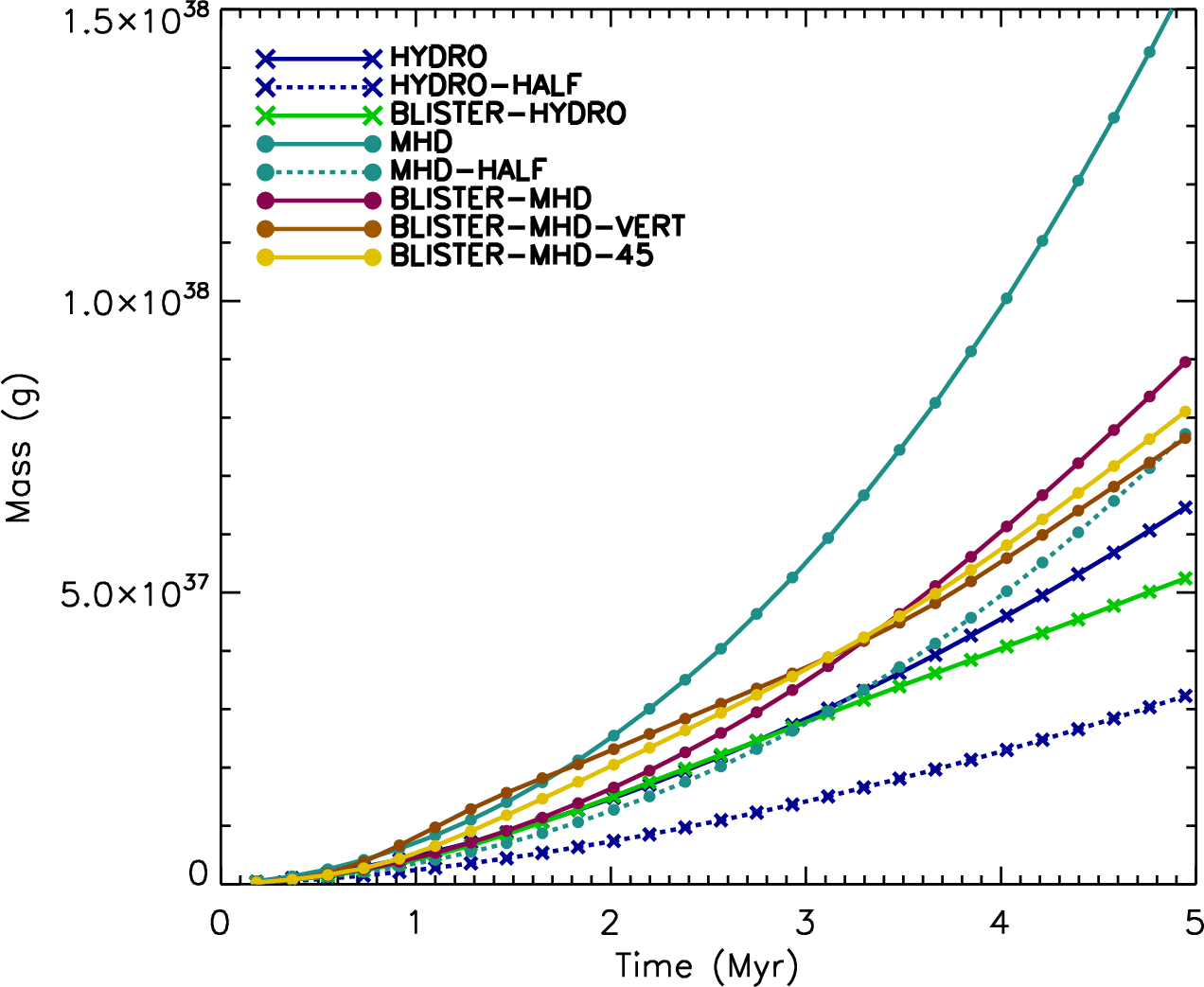}
\caption{The mass of the various runs summed over cells whose density is greater than 1.01 times the initial density. The dashed curves represent the hydro and mhd curves divided in half for comparison.}
\label{fig:mass-plot} 
\end{figure}

We also calculate how much mass is swept up in the dense shell (as well as in the fast magnetosonic wave) and present these results in Fig. \ref{fig:mass-plot}.  The dashed lines represent the hydro and mhd runs divided in half. Although the symmetric runs sweep up more mass than their blister counterparts, the difference is less than a factor of two, so the fact that the blister runs send hemispherical rather than a spherical shell into the cloud is partly compensated for by the fact that the shell expands more rapidly. The blister-mhd run sweeps up nearly twice as much mass by 5 Myr as the blister-hydro run, so the magnetic fields make a large difference with respect to sweeping up of mass. The orientation of the initial magnetic field makes little difference in terms of sweeping up of mass for the blister-type runs. 

\subsubsection{Kinetic Energy}
\label{sec:kinetic-energy}

\begin{figure}
\centering
\includegraphics[scale=0.19]{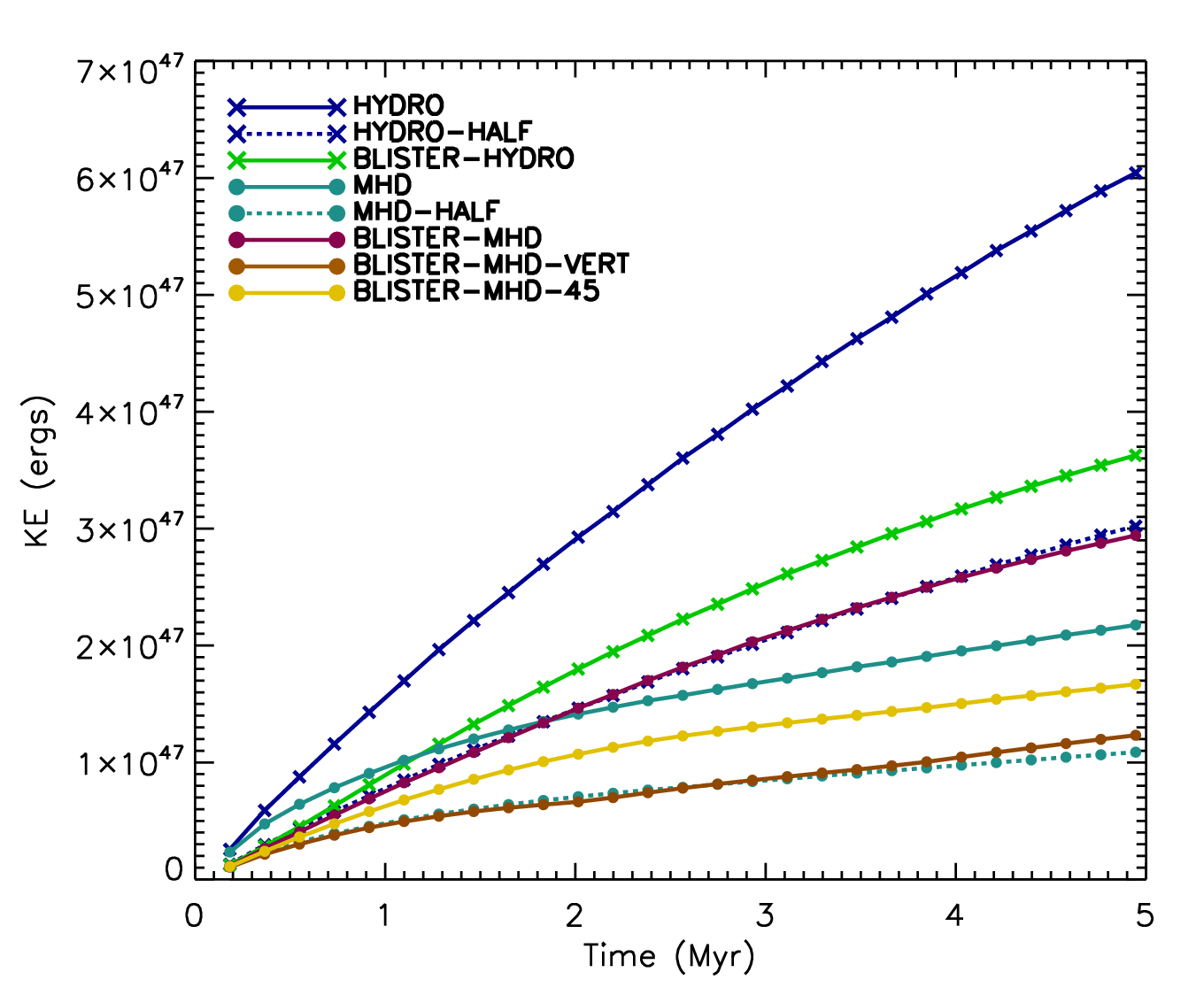}
\caption{The total kinetic energy of all the runs. The kinetic energy is calculated only considering cells whose density is greater than 1.01 times the initial density. The dashed curves represent the hydro and mhd curves divided in half.}
\label{fig:kinetic-energy-comparison}
\end{figure}

\begin{figure}
\centering
\includegraphics[scale=0.19]{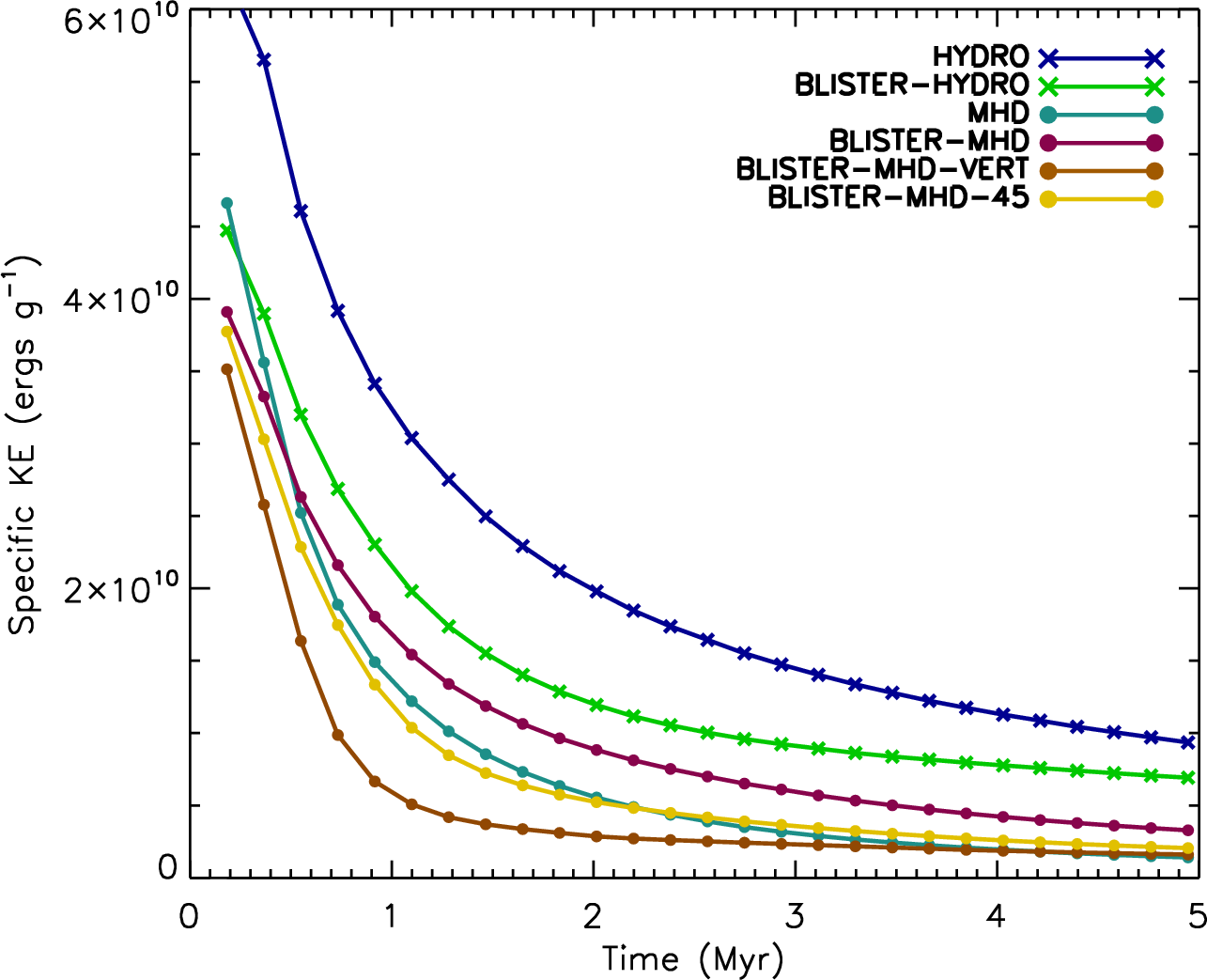}
\caption{The specific kinetic energy of all the runs calculated only considering cells whose density is greater than 1.01 times the initial density.}
\label{fig:kinetic-energy-overmass-comparison}
\end{figure}

We plot the kinetic energies of the various runs in Fig. \ref{fig:kinetic-energy-comparison} and the specific kinetic energy, defined as the kinetic energy divided by the swept up mass, in Fig. \ref{fig:kinetic-energy-overmass-comparison}. Note that, since we only include cells with $\rho>1.01 \rho_0$ in this computation, we are only evaluating the kinetic energy imparted to the dense cloud, and not the kinetic energy carried by the outflowing gas or deposited in the low density medium.

The result that the blister-hydro kinetic energy is lower than the hydro kinetic energy is contrary to what we would expect to see based on the analytic solution. From Eqn's. \eqref{Eqn:Spitzer1978-A} and \eqref{Eqn:Spitzer1978-B}, we see that the predicted blister radius is larger by a factor of $2^{2/7}$ than the symmetric radius for any fixed time. The kinetic energy of the shell is given by $E_{KE}=1/2 M\dot{r}_{sh}^{2}$, where M is the mass of the shell. The mass of the shell increases as $r_{sh}^{3}$, so $E_{KE}$ should be larger by a factor of $(1/2)(2^{6/7})(2^{2/7})^{2}=2^{3/7}$ for the blister-hydro run, where the $1/2$ term is included to account for the fact that the blister-hydro shell is a hemisphere rather than a sphere. However, the spherical part of the blister-hydro shell actually expands significantly slower than the analytic solution predicts (see \S \ref{sec:discussion}), so if most of the kinetic energy is concentrated in the spherical part of the shell, the total kinetic energy for the blister-hydro run could be lower than for the hydro run. This is indeed the case. As seen in Fig. \ref{fig:blister-hydro} row 2, the slivers of the shell that expand in the $\hat{y}$ direction have very little kinetic energy density. Thus, even though they contribute significantly to the radius of the blister-hydro shell, the total kinetic energy is lower in the blister-hydro case than in the hydro case. 

In general, the non-mhd runs have both more total and specific kinetic energy than their magnetic counterparts. The total and specific kinetic energy is very sensitive to the orientation of the initial magnetic field. The blister-mhd run has more total and specific kinetic energy than the mhd run, but the mhd run has more total kinetic energy and a comparable amount of specific kinetic energy to the blister-mhd-vert and blister-mhd-45 runs. It is interesting to compare the effect of going from the mhd to the blister-mhd run to that of going from the hydro to the blister-hydro run. Going from symmetric to blister has the effect of increasing both the total and specific kinetic energies in the presence of a magnetic field, but has the opposite effect in their absence. It is also interesting to note that the blister-hydro run has less specific kinetic energy than the hydro run. This is not surprising, since the mass swept up by the blister-hydro run is significantly larger than half of the mass swept up in the hydro case (Fig. \ref{fig:mass-plot}).

\subsubsection{Magnetic Energy}
\label{sec:magnetic-energy}

\begin{figure}
\centering
\includegraphics[scale=0.19]{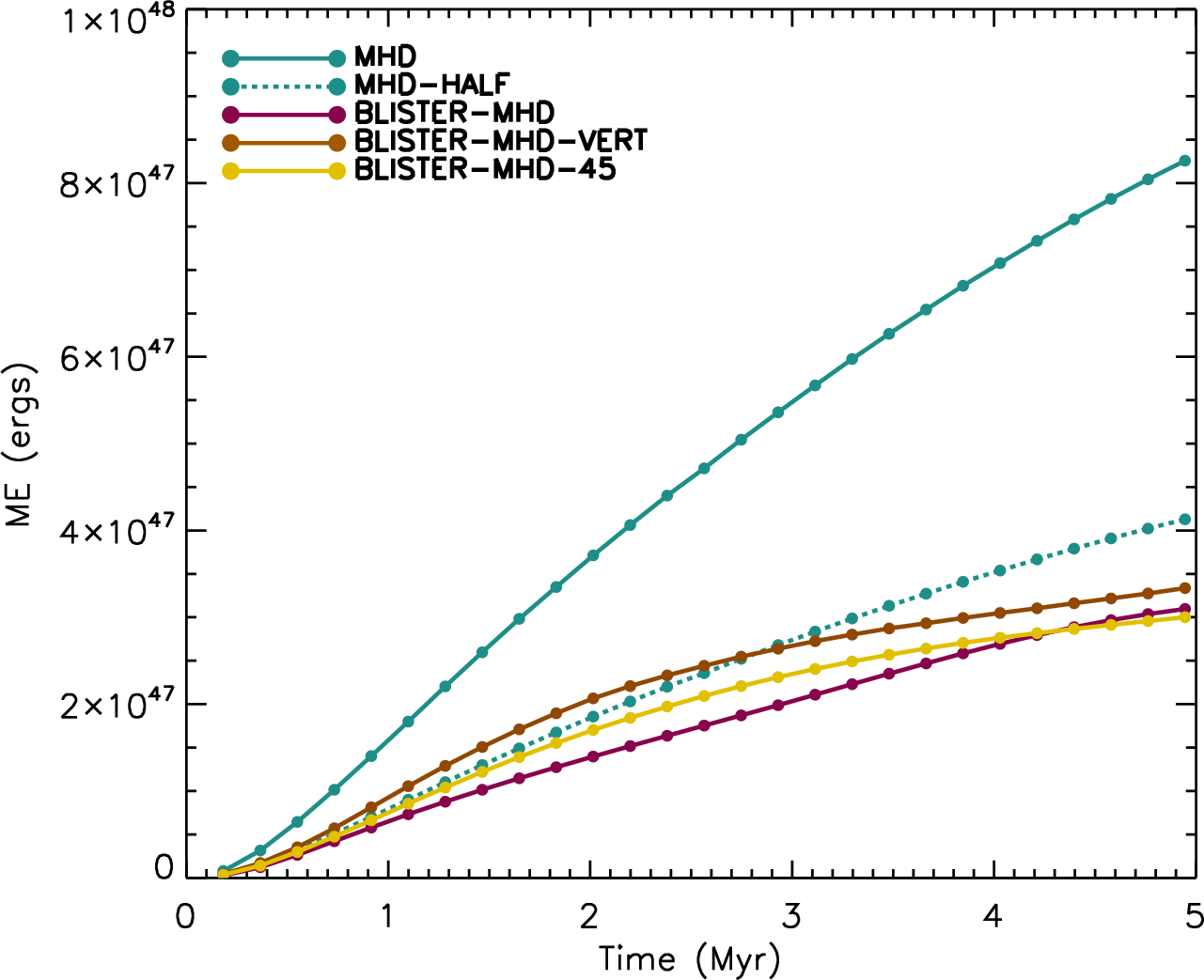}
\caption{The change in total magnetic energy of all the runs. The magnetic energy is calculated only considering cells whose density is greater than 1.01 times the initial density. The dashed curve represents the mhd curve divided in half.}
\label{fig:magnetic-energy-comparison}
\end{figure}

For the magnetic energy, we expect the results to be opposite of those for the kinetic energy. The runs with the least kinetic energy should actually have the most magnetic energy since energy that does not go into motion is instead stored as distortions of the magnetic field. We present these results in Fig. \ref{fig:magnetic-energy-comparison}. The dashed line is the mhd run divided in half for comparison. 

We calculate the change in magnetic energy using the same criterion as for the kinetic energy ($\rho > 1.01 \rho_{0}$) not only to avoid taking into account fluctuations that occur outside of the cloud, but also to exclude cells on the edge of the grid. In our simulations the fast magnetosonic wave eventually reaches the edge of the grid, stretching the magnetic field lines outside of the computational domain. If we were to take into account edge cells, we could see a net loss of magnetic energy as the fast magnetosonic wave leaves the grid, even though in reality magnetic energy is constantly being injected into the GMC.

\begin{figure}
\centering
\includegraphics[scale=0.19]{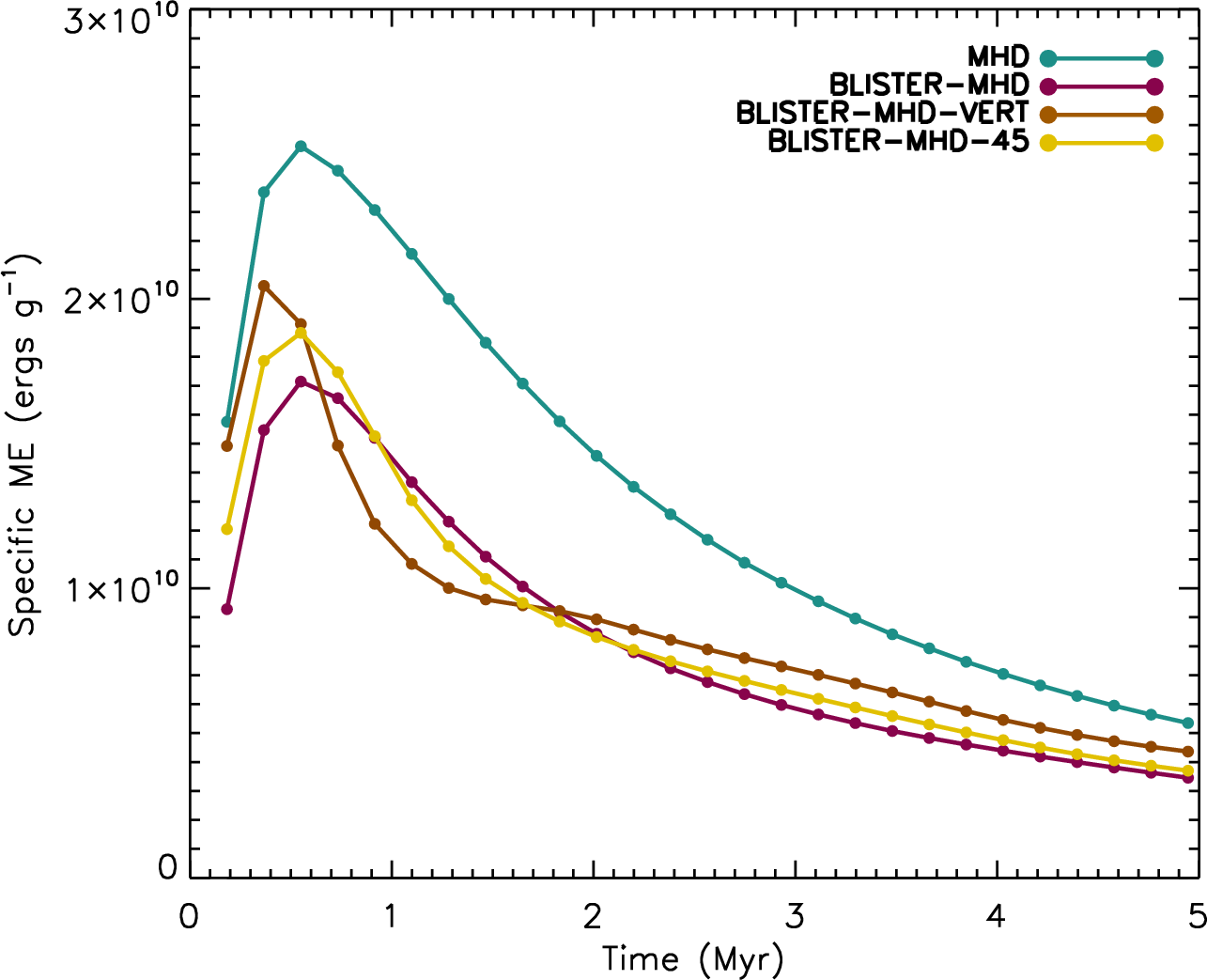}
\caption{The change in specific magnetic energy.}
\label{fig:magnetic-energy-overmass-comparison}
\end{figure}

In order to account for only the magnetic energy injected into the GMC by the HII region and not the initial magnetic energy present in each cell, we calculate the total change in magnetic energy, $\Delta E_{B}$, where $\Delta E_{B} = ((B_{x})^{2}+(B_{y})^{2}+(B_{z})^{2})/2 - ((B_{x_{0}})^{2}+(B_{y_{0}})^{2}+(B_{z_{0}})^{2})/2$, where $B_{x_{0}}, B_{y_{0}}, B_{z_{0}}$ is the initial magnetic field strength in the x, y, and z directions, respectively. We find that this quantity is very similar for all the blister runs. Fig. \ref{fig:magnetic-energy-overmass-comparison}  shows the specific change in magnetic energy $\Delta E_{B}/M$. The curves all decline with time at late times, which implies that the rate at which the magnetic energy is injected is slower than the rate at which mass is swept up by the ionization front for all the runs. The symmetric HII regions have both more total magnetic energy and specific magnetic energy than their blister counterparts, although at late times the difference in the specific magnetic energy is very small. 

\subsubsection{Total Energy}
\label{sec:total-energy}

\begin{figure}
\centering
\includegraphics[scale=0.19]{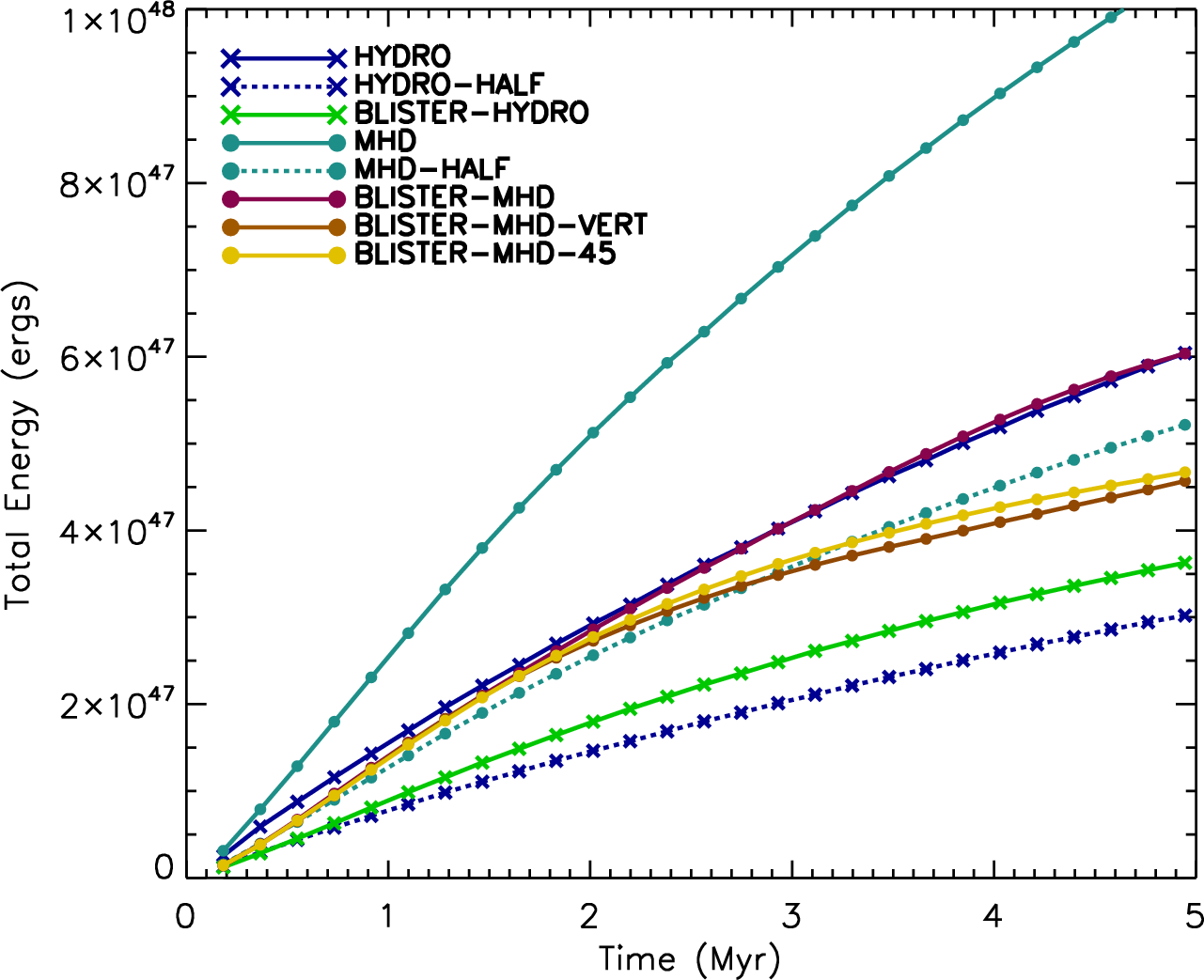}
\caption{The total energy of all the runs. The dashed curves represent the hydro and MHD runs divided by a factor of two.}
\label{fig:total-energy-comparison}
\end{figure}

\begin{figure}
\centering
\includegraphics[scale=0.19]{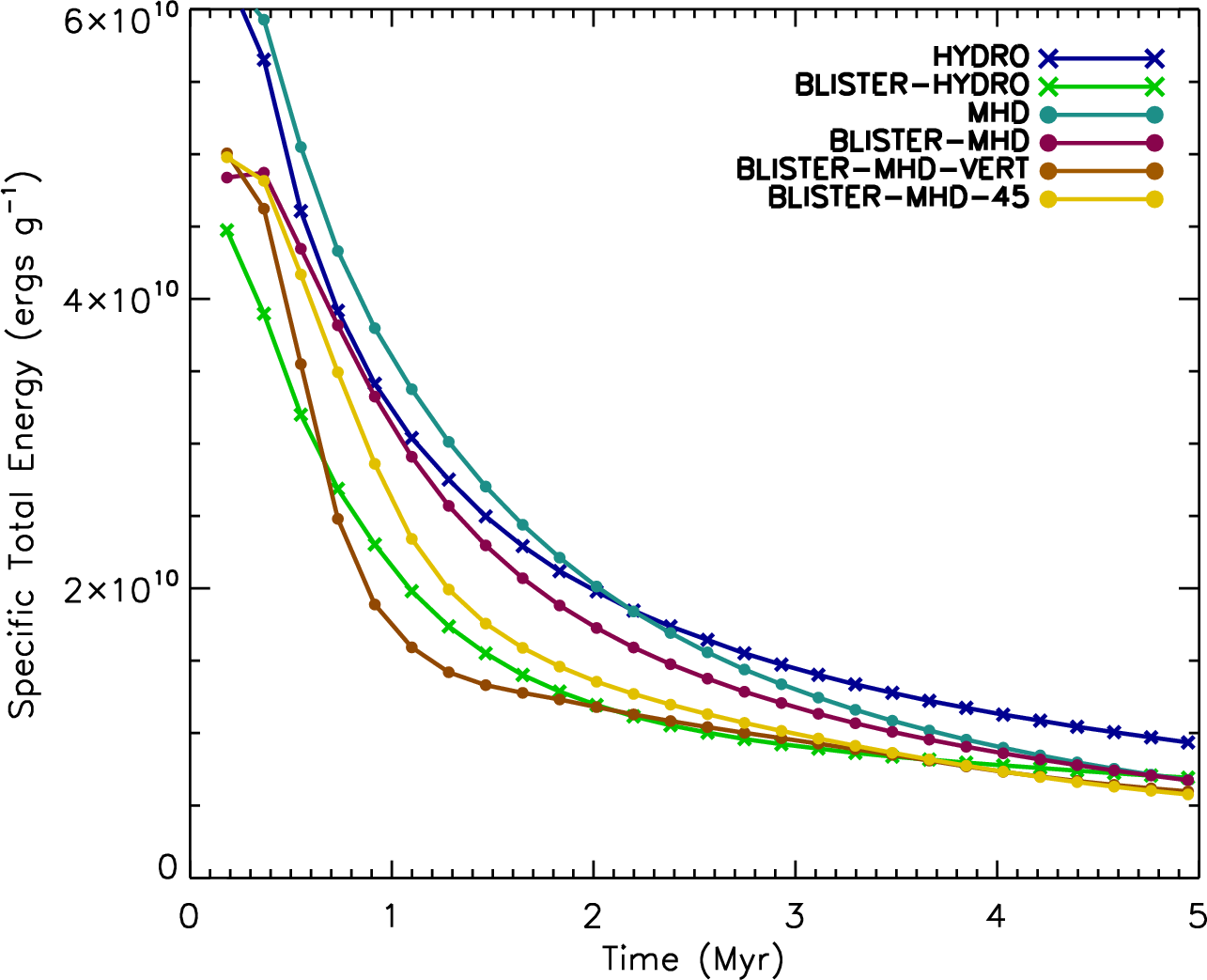}
\caption{The specific total energy of all the runs.}
\label{fig:total-energy-overmass-comparison}
\end{figure}

In Fig. \ref{fig:total-energy-comparison}, we plot the total energy, that is, kinetic plus magnetic energy, for all the runs. For the hydro and blister-hydro runs, there is no magnetic energy of course, so this just consists entirely of the kinetic energy. It is clear that magnetic effects are of great importance with respect to the total energy of the cloud. By 5 Myr, the blister-mhd run has about twice as much energy as the blister-hydro run, and the blister-mhd-vert and blister-mhd-45 runs have about 30 $\%$ more energy. We plot the specific total energy in Fig. \ref{fig:total-energy-overmass-comparison}. The specific total energy depends most strongly on the magnetic field orientation and not on whether it is a blister or symmetric case. There is no advantage in going from hydro to mhd in the symmetric case, but there is an advantage in doing so for the blister case at least for some magnetic field orientations. 

It is instructive to compare the kinetic energy lost to the magnetic energy gained in going from blister-hydro to blister-mhd. The blister-mhd case has $\approx 6\times 10^{46}$ erg less kinetic energy at 5 Myr, but it also has $\approx 3\times 10^{47}$ erg more magnetic energy. Thus the gain in magnetic energy outweighs the loss of kinetic energy by a factor of 5. As a result, we find that in general that magnetized HII regions deliver significantly more energy to dense clouds than their pure hydrodynamic counterparts.

\section{Discussion}

\label{sec:discussion}
Here we investigate the origin of the differences between the analytic approximations and the numerical results that we described in the previous section. 

The validity of the approximations used to derive the analytic approximation can be checked through a simple calculation. Using Eqn. \eqref{eqn:Stromgren-Radius} to estimate the HII region internal density, and a value for the ionizing luminosity from \S \ref{sec:problem-setup}, we find that the ratio of ambient pressure to HII region pressure is $\approx 4.4 \%$ at 1 Myr and $\approx 12.3 \%$ at 3 Myr. Thus although the approximation of HII region pressure dominance works quite well at early times, by 3 Myr it is not quite as accurate, and this discrepancy can cause some of the observed flattening of the blister-hydro and spherical-hydro curves late in the run, although clearly not enough to account for the substational flattening in the blister-hydro run observed in Fig. \ref{fig:radius-comparison}.

\begin{figure}
\centering
\includegraphics[scale=0.19]{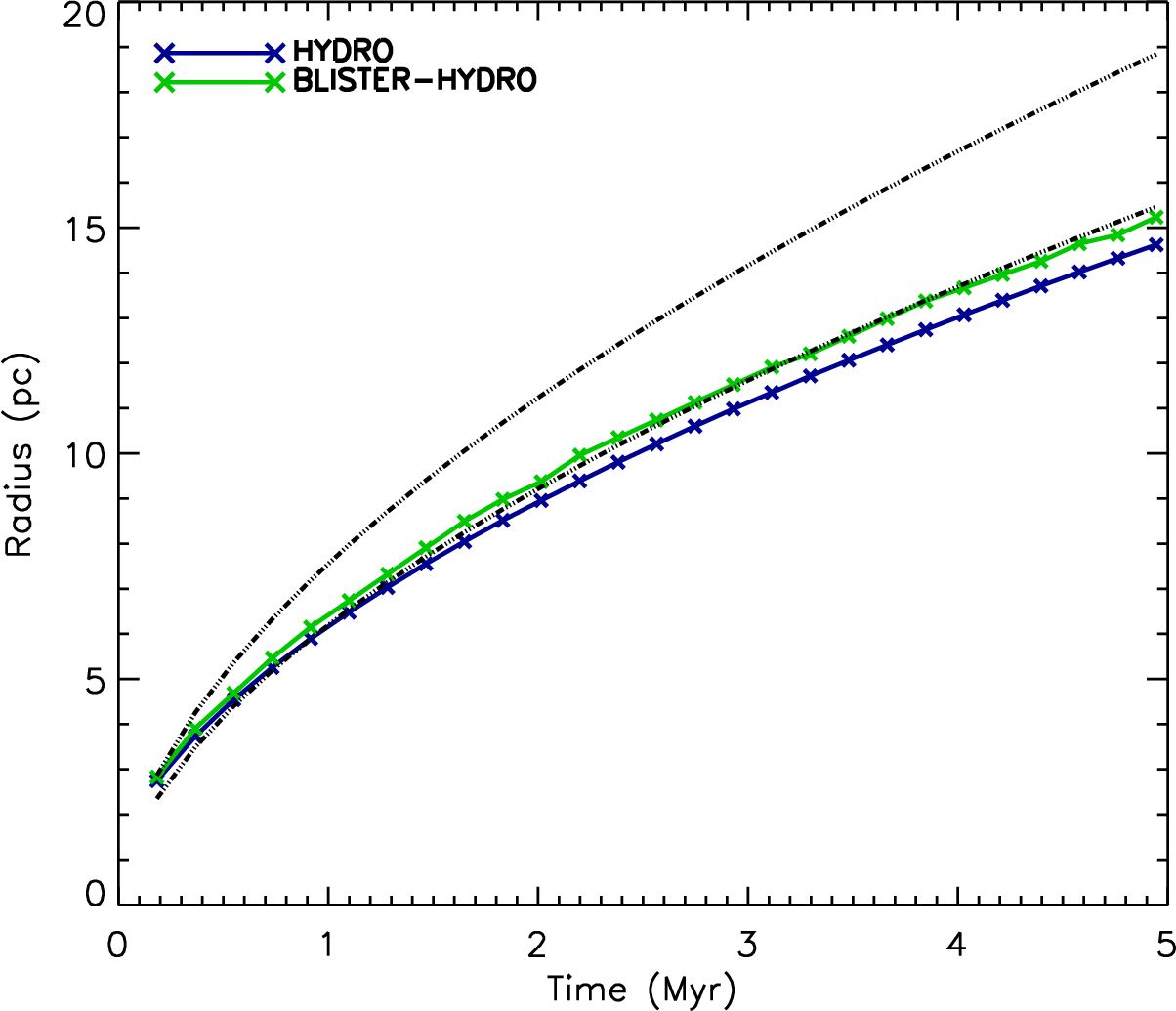}
\caption{The radius for the blister-hydro run computed by only considering gas along the $y=z=0$ line. The blue-dashed curves are the same as in Fig. \ref{fig:radius-comparison}.}
\label{fig:blister-hydro-line-plot}
\end{figure}

In reality this flattening is probably due to the definition used to calculate the radius of the blister-hydro curve. The radius is a bit misleading because it turns out that the spherical portion of the blister-hydro shell travels significantly slower than the analytic solution predicts. This is shown in Fig. \ref{fig:blister-hydro-line-plot}, where we have calculated the radius of the blister-hydro shell by only considering gas along the $y=z=0$  line. The spherical part of the shell expands just a bit faster than the hydro shell. This implies that the slivers of the shell that extend in the $\hat{y}$ direction along the interface (as seen in Fig. \ref{fig:blister-hydro}) contribute significantly to the radius of the blister hydro curve. In the early to mid-stages of the simulation the slivers expand faster in the y-direction than the spherical part of the shell expands in the radial direction, but towards the later stages the expansion of the slivers in the y-direction slows down dramatically and the curve becomes flatter than the analytic solution. 

\begin{figure}
\centering
\includegraphics[scale=0.185]{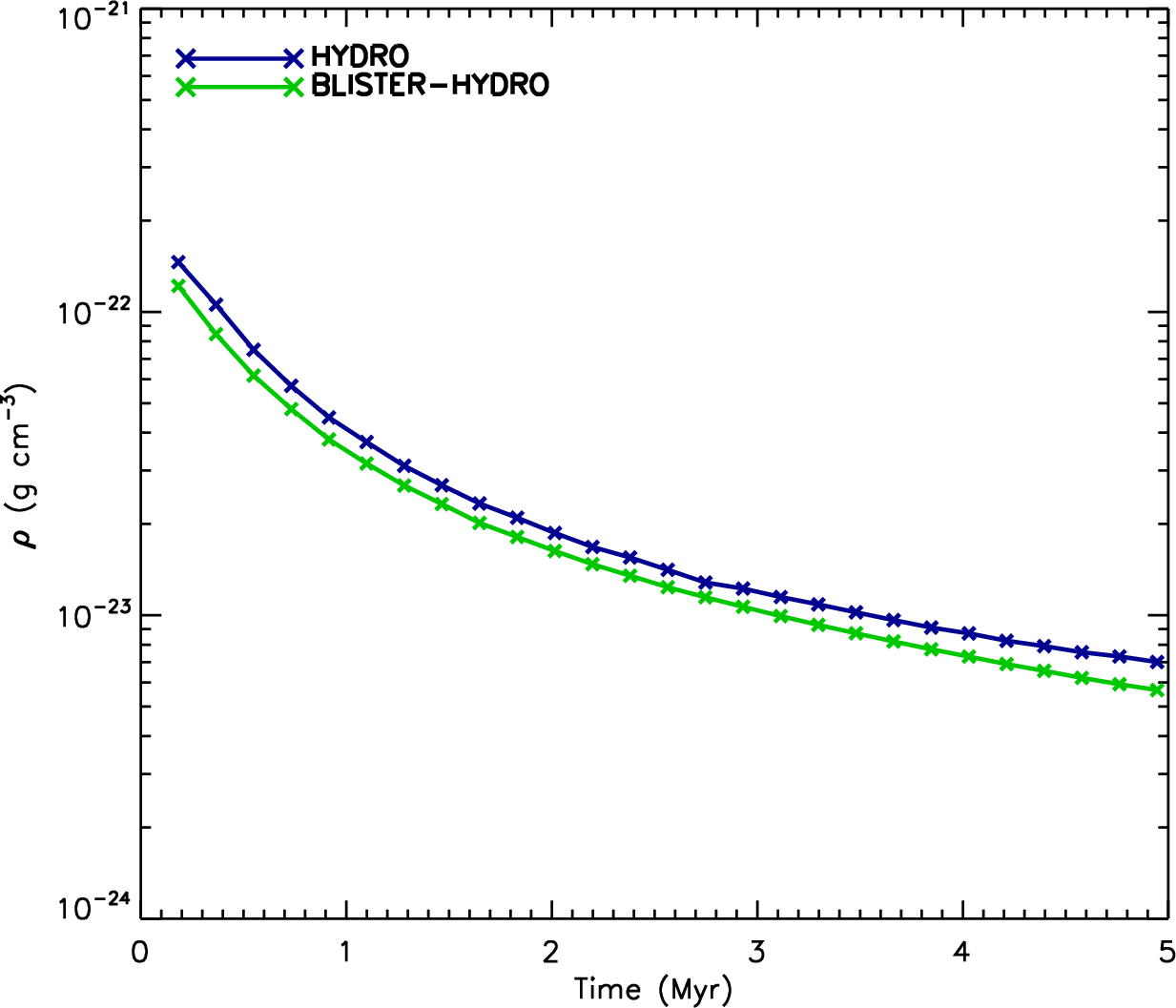}
\caption{The density just inside the dense shell of the hydro and blister-hydro HII regions plotted over time.}
\label{fig:hii-region-density}
\end{figure}

While this may explain the shape of the curve, it does not explain why the blister-hydro shell seems to expand at a rate comparable to the hydro shell, rather than roughly 20\% faster as predicted by the analytic approximation. 
One possible explanation for this discrepancy is that the analytic solution for the blister case assumes that the density is the same as in the symmetric HII region. The D-type ionization front travels below the sound speed of the HII region, so gas inside the symmetric HII region has time to spread out and achieve a uniform density. This is not quite true for the blister case since the gas is free to escape from the HII region, so that there is a non-uniform density distribution inside the blister-type HII region. We check this by comparing the density just inside the shell for both the blister-hydro and hydro runs (Fig. \ref{fig:hii-region-density}). The blister-hydro density is about 15\% less than the hydro density for the entire run. If we multiply $\rho_{II}$ in Eqn. \ref{eqn:ionization-balance} by a factor of 0.85 in the blister case to account for this effect, we find the predicted difference in shell radius and speed between the symmetric and blister cases drops from 20\% to 15\%. Thus the incorrect assumption of a uniform density in the blister case accounts for about a quarter of the discrepancy. The rest is likely due to a failure of the assumption of hemispherical symmetry. Comparing Fig.'s \ref{fig:radius-comparison} and \ref{fig:blister-hydro-line-plot}, we see that the mass-averaged radius considering all angles is $\sim 50\%$ larger than the value along the $y=z=0$ line even at very early times, before significant tails form. Thus the shell is only very roughly hemispherical. These results show that the blister HII regions are much more complex than their symmetric counterparts, and that the analytic solution for the blister case cannot use the same simplifying assumptions that work quite well for the symmetric case.

\section{Conclusions}

\label{section:conclusion}

We have performed the first numerical study of a blister type HII region expanding into a magnetized medium. We draw the following conclusions:

(i)  Although the kinetic energy of the magnetized runs is lower than that of their hydrodynamic counterparts, they have much more total energy since the kinetic energy lost in going from hydro to MHD is many times less than the magnetic energy gained (\S \ref{sec:total-energy}), and hence could possibly be more efficient at driving turbulence. It is not entirely clear how efficient the injected magnetic energy is at driving turbulence compared to the kinetic energy, but studies of Alfven waves decaying into turbulence in various astrophysical environments (from GMCs to the solar wind) suggest that the process is likely to be very efficient. A circular Alfven wave develops a ``decay'' instability which ultimately leads to decay into turbulence (see \citet{McKee2007} and references therein). This decay into turbulence requires an initial directional imbalance in the Alfven waves, which we have in our simulations since the waves are all left-propogating. It is likely that the magnetic energy added to the cloud by including the effects of MHD is at least as important, if not more important, than the kinetic energy for any type of HII region. Therefore it is important to include the effects of MHD in future studies of star formation. 

(ii) A blister-type HII region expands into a cloud faster than the corresponding symmetric case, but not by as much as predicted by some analytic approximations. 

(iii) The total energy is greatest in the symmetric case, so HII regions of this type make the greatest contribution to the total energy budget of a cloud. However, since GMCs are turbulent and have a filamentary morphology, most new born stars are likely to be near the edge, so the blister scenario should be more common than the embedded one. Nonetheless, our simulations confirm that, in the presence of a magnetic field, even blister-type HII regions can inject significant energy into the dense parts of molecular clouds.

\acknowledgements
We thank J.~Stone, T.~Gardiner, and P.~Teuben for writing the Athena MHD code. This work was supported by the Alfred P.\ Sloan Foundation, the NSF through grants AST-0807739 and CAREER-0955300, and NASA through ATFP grant NNX09AK31G, a Spitzer Space Telescope theoretical research program grant, and a Chandra Space Telescope grant.

\begin{appendix}
\section{Derivation of the Shell Expansion Rate}
\label{sec:appendix}
Here we derive the analytic solution of the radius of the shell in the spherical and blister cases. We assume ionization balance, consider the density $\rho_{II}$ constant inside the HII region, and use momentum conservation: $\frac{d\mathcal{P}}{dt} = \mathcal{F}$, where $\mathcal{P}$ is the momentum of the shell and $\mathcal{F}$ is the force applied to it by the matter inside the HII region. The mass of the shell is $M_{sh}=(4,2)\pi r^{3}\rho_{0}/3$ since most of the mass inside the Stromgren Sphere is contained in the shell to good approximation, where 4 and 2 are the coefficients for the symmetric and blister cases, respectively. Using ionization balance and Eqn. \eqref{eqn:Stromgren-Radius} we can write the density inside the HII region
\begin{equation}
\label{eqn:ionization-balance}
\rho_{II} = \left( \frac{3s\mu_{H}^{2}}{4\pi \alpha^{(B)}} \right)^{1/2}r^{-3/2},
\end{equation}
and hence the pressure (thermal plus ram) inside the HII region
\begin{equation}
P = (1,2)\rho_{II}c_{II}^{2}=(1,2)c_{II}^{2}\left( \frac{3s\mu_{H}^{2}}{4\pi \alpha^{(B)}} \right)^{1/2}r^{-3/2},
\end{equation}
where the coefficient of 2 for the blister case represents the ram pressure of material rocketing off the inside of the dense shell as it is ionized \citep{Matzner2009}. Now we use momentum conservation to arrive at the equation of motion:
\begin{equation}
\frac{d\mathcal{P}}{dt} = \frac{d}{dt}\left[(4,2)\pi r^3\rho_{0} \dot{r}/3 \right] = \mathcal{F} = P \mathcal{A} =
4\pi r^{2}c_{II}^{2}\left( \frac{3s\mu_{H}^{2}}{4\pi \alpha^{(B)}} \right)^{1/2}r^{-3/2} = 4 c_{II}^{2} \left( \frac{3s\mu_{H}^{2}}{4\pi \alpha^{(B)}} \right)^{1/2} r^{1/2},
\end{equation}
where $\mathcal{A}$ is the surface area of the shell. 
\begin{equation}
\label{eqn:equation-of-motion}
\Longrightarrow \frac{(4,2)\rho_{0}}{3}\left[r^{3}\ddot{r}+3r^{2}\dot{r}^{2}\right] = 4 c_{II}^{2}\left( \frac{3s\mu_{H}^{2}}{4\pi \alpha^{(B)}} \right)^{1/2}r^{1/2}.
\end{equation}
This ODE admits a similarity solution of the form $r\propto t^{\eta}$, and with some algebra one can show that 
\begin{equation}
r_{sh}=r_{s}\left ( \frac{7t}{\sqrt{12} t_{s}} \right )^{4/7}       \mbox{(spherical)},
\end{equation}
and
\begin{equation}
r_{sh}=r_{s}\left ( \frac{7t}{\sqrt6 t_{s}} \right )^{4/7}       \mbox{(blister)}, 
\end{equation}
where $t_s = r_s/c_{II}$.
\end{appendix}

\bibliographystyle{apj}
\bibliography{thesis_refs}

\end{document}